\documentclass[a4paper,12pt]{article}
\usepackage{epsfig}
\usepackage{amssymb}
\usepackage{amsfonts}
\usepackage{amsmath}
\usepackage{euscript}
\usepackage{verbatim}
\usepackage{latexsym}
\usepackage{graphicx}
\usepackage{caption, color}
\usepackage{float}
\usepackage{cite, hyperref}
\usepackage{subcaption}

\usepackage[all]{xy}

\hypersetup{ linktoc=all,
    colorlinks, linkcolor={palatinateblue},
    citecolor={brightpink}, urlcolor={amaranth}
}

\graphicspath{{Images/}}

\definecolor{rosy}{RGB}{230,235,252}
\definecolor{myframetitle}{RGB}{90,89,170}
\definecolor{myblocktitle}{RGB}{140,185,249}
\definecolor{mytitle}{RGB}{10,80,26}

\definecolor{darkgreen}{RGB}{27,130,45}
\definecolor{darkblue}{rgb}{0,0,0.3}
\definecolor{darkred}{rgb}{0.7,0,0}

\definecolor{light gray}{RGB}{220,220,220}
\definecolor{dark purple}{RGB}{108,0,217}
\definecolor{pink}{RGB}{190,20,100}
\definecolor{orang}{RGB}{193,63,0}
\definecolor{green}{RGB}{11,98,17}
\definecolor{darkpink}{RGB}{153,0,76}
\definecolor{bluegreen}{RGB}{0,102,102}
\definecolor{greenlagan}{RGB}{0,102,0}
\definecolor{redgreen}{RGB}{102,102,0}
\definecolor{Redgreen}{RGB}{153,76,0}
\definecolor{vividviolet}{rgb}{0.62, 0.0, 1.0}
\definecolor{amaranth}{rgb}{0.9, 0.17, 0.31}
\definecolor{palatinateblue}{rgb}{0.15, 0.23, 0.89}
\definecolor{brightpink}{rgb}{1.0, 0.0, 0.5}
\definecolor{cornflowerblue}{rgb}{0.39, 0.58, 0.93}
\definecolor{deepcarminepink}{rgb}{0.94, 0.19, 0.22}
\definecolor{radicalred}{rgb}{1.0, 0.21, 0.37}

\usepackage{listings}

\newif\ifdtup

\def\lcdm{$\Lambda$CDM}

\catcode`\@=11

\@addtoreset{equation}{section}

\def\@normalsize{\@setsize\normalsize{15pt}\xiipt\@xiipt
\abovedisplayskip 14pt plus3pt minus3pt%
\belowdisplayskip \abovedisplayskip
\abovedisplayshortskip \z@ plus3pt%
\belowdisplayshortskip 7pt plus3.5pt minus0pt}

\def\small{\@setsize\small{13.6pt}\xipt\@xipt
\abovedisplayskip 13pt plus3pt minus3pt%
\belowdisplayskip \abovedisplayskip
\abovedisplayshortskip \z@ plus3pt%
\belowdisplayshortskip 7pt plus3.5pt minus0pt
\def\@listi{\parsep 4.5pt plus 2pt minus 1pt
     \itemsep \parsep
     \topsep 9pt plus 3pt minus 3pt}}

\relax

\catcode`@=12

\catcode`\@=11

\def\section{\@startsection{section}{1}{\z@}{3.5ex plus 1ex minus
   .2ex}{2.3ex plus .2ex}{\large\bf}}

\def\SymBoxes#1#2#3#4{\newdimen\un@t \un@t#3%
\raisebox{#1}{\rule{#2\un@t}{#4}\hskip-#2\un@t
\@tempdimb\un@t \advance\@tempdimb by-#4\@tempcntb#2\relax%
\@whilenum{\@tempcntb>0}\do{
\rule{#4}{\un@t}\hskip\@tempdimb \advance\@tempcntb by\m@ne}%
\hskip-#2\un@t \rule[\un@t]{#2\un@t}{#4}%
\rule[\un@t]{#4}{#4}\hskip-#4
\rule{#4}{\un@t}}\hskip-#4}                

\begin{document}

\newcommand{\beq}{\begin{equation}}
\newcommand{\eeq}{\end{equation}}
\newcommand{\bea}{\begin{eqnarray}}
\newcommand{\eea}{\end{eqnarray}}
\newcommand{\beas}{\begin{eqnarray*}}
\newcommand{\eeas}{\end{eqnarray*}}
\newcommand{\defi}{\stackrel{\rm def}{=}}
\newcommand{\non}{\nonumber}
\newcommand{\bquo}{\begin{quote}}
\newcommand{\enqu}{\end{quote}}
\renewcommand{\(}{\begin{equation}}
\renewcommand{\)}{\end{equation}}
\def \eqn#1#2{\begin{equation}#2\label{#1}\end{equation}}

\def\e{\epsilon}
\def\IZ{{\mathbb Z}}
\def\IR{{\mathbb R}}
\def\IC{{\mathbb C}}
\def\IQ{{\mathbb Q}}
\def\de{\partial}
\def\Tr{ \hbox{\rm Tr}}
\def\H{ \hbox{\rm H}}
\def\HE{ \hbox{$\rm H^{even}$}}
\def\HO{ \hbox{$\rm H^{odd}$}}
\def\K{ \hbox{\rm K}}
\def\Im{ \hbox{\rm Im}}
\def\Ker{ \hbox{\rm Ker}}
\def\const{\hbox {\rm const.}}
\def\o{\over}
\def\im{\hbox{\rm Im}}
\def\re{\hbox{\rm Re}}
\def\bra{\langle}\def\ket{\rangle}
\def\Arg{\hbox {\rm Arg}}
\def\Re{\hbox {\rm Re}}
\def\Im{\hbox {\rm Im}}
\def\exo{\hbox {\rm exp}}
\def\diag{\hbox{\rm diag}}
\def\longvert{{\rule[-2mm]{0.1mm}{7mm}}\,}
\def\a{\alpha}
\def\dag{{}^{\dagger}}
\def\tq{{\widetilde q}}
\def\p{{}^{\prime}}
\def\W{W}
\def\N{{\cal N}}
\def\hsp{,\hspace{.7cm}}

\def\br{\nonumber}
\def\IZ{{\mathbb Z}}
\def\IR{{\mathbb R}}
\def\IC{{\mathbb C}}
\def\IQ{{\mathbb Q}}
\def\IP{{\mathbb P}}
\def \eqn#1#2{\begin{equation}#2\label{#1}\end{equation}}

\newcommand{\C}{\ensuremath{\mathbb C}}
\newcommand{\Z}{\ensuremath{\mathbb Z}}
\newcommand{\R}{\ensuremath{\mathbb R}}
\newcommand{\rp}{\ensuremath{\mathbb {RP}}}
\newcommand{\cp}{\ensuremath{\mathbb {CP}}}
\newcommand{\vac}{\ensuremath{|0\rangle}}
\newcommand{\vact}{\ensuremath{|00\rangle}                    }
\newcommand{\oc}{\ensuremath{\overline{c}}}
\newcommand{\psizero}{\psi_{0}}
\newcommand{\phizero}{\phi_{0}}
\newcommand{\hzero}{h_{0}}
\newcommand{\psiin}{\psi_{\rh}}
\newcommand{\phiin}{\phi_{\rh}}
\newcommand{\hin}{h_{\rh}}
\newcommand{\rh}{r_{h}}
\newcommand{\rb}{r_{b}}
\newcommand{\psibnd}{\psi_{0}^{b}}
\newcommand{\psibndp}{\psi_{1}^{b}}
\newcommand{\phibnd}{\phi_{0}^{b}}
\newcommand{\phibndp}{\phi_{1}^{b}}
\newcommand{\gbnd}{g_{0}^{b}}
\newcommand{\hbnd}{h_{0}^{b}}
\newcommand{\zh}{z_{h}}
\newcommand{\zb}{z_{b}}
\newcommand{\man}{\mathcal{M}}
\newcommand{\hbr}{\bar{h}}
\newcommand{\tbr}{\bar{t}}

\newcommand\tcr{\textcolor{red}}
\newcommand\tcb{\textcolor{blue}}
\newcommand\tcg{\textcolor{green}}

\newcommand\snote[1]{\textcolor{red}{\bf [Sh:\,#1]}}

\begin{titlepage}
\begin{flushright}
CHEP XXXXX \\
IPM/P-2022/nnnn
\end{flushright}
\bigskip
\def\thefootnote{\fnsymbol{footnote}}

\begin{center}
\centerline{\Large{\bf Dipole Cosmology: The Copernican Paradigm Beyond FLRW}}
\end{center}

\bigskip
\begin{center}
Chethan KRISHNAN$^a$\footnote{\texttt{chethan.krishnan@gmail.com}}, \ \ Ranjini MONDOL$^a$\footnote{\texttt{ranjinim@iisc.ac.in}}, \ \ M. M. SHEIKH-JABBARI$^b$\footnote{\texttt{jabbari@theory.ipm.ac.ir}}
\vspace{0.1in}

\end{center}

\renewcommand{\thefootnote}{\arabic{footnote}}

\begin{center}

$^a$ {Center for High Energy Physics,\\
Indian Institute of Science, Bangalore 560012, India}\\

{$^b$ School of Physics, Institute for Research in Fundamental
Sciences (IPM),\\ P. O. Box 19395-5531, Tehran, Iran}

\end{center}

\noindent
\begin{center} {\bf Abstract} \end{center}
We introduce the \emph{dipole cosmological principle}, the idea that the Universe is a maximally Copernican cosmology,  compatible with a cosmic flow. It serves as the most symmetric paradigm  that generalizes the FLRW ansatz, in light of the increasingly numerous (but still tentative) hints that have emerged in the last two decades for a non-kinematic component in the CMB dipole. Einstein equations in our ``dipole cosmology" are still ordinary differential equations -- but instead of the two Friedmann equations, now we have four.  The two new functions can be viewed as an anisotropic scale factor that breaks the isotropy group from $SO(3)$ to $U(1)$, and a ``tilt'' that captures the cosmic flow velocity. The result is an axially isotropic, tilted Bianchi V/VII$_h$ cosmology. We assess the possibility of model building within the dipole cosmology paradigm, and discuss the dynamics of expansion rate, anisotropic shear and tilt, in various examples. A key observation is that the cosmic flow (tilt) can grow even while the anisotropy (shear) dies down. Remarkably, this can happen even in an era of late time acceleration.

\vspace{1.6 cm}
\vfill

\end{titlepage}

\setcounter{footnote}{0}

\tableofcontents 

\section{Introduction and Motivation, A Breakdown of FLRW Setup?}

The philosophy that we are non-privileged observers has been a powerful one in the history of science, ever since Copernicus \cite{Copernicus}. In modern cosmology, the idea that our location is not special in time nor space is sometimes called the Perfect Cosmological Principle. 
A maximally symmetric spacetime like Minkowski space or (anti-)de Sitter space would satisfy this prejudice and would be a candidate for the ultimate Copernican cosmology, but it suffers from a serious problem: it is wrong. 
Hubble's discovery of the expansion of the Universe as well as the enormous evidence for the Big Bang paradigm, 
tell us that time translation is not a symmetry of our background on large scales and that maximally symmetric spacetimes cannot describe our Universe.\footnote{There is a slight caveat here regarding de Sitter space which is in fact maximally symmetric, but has time-dependence in the cosmological patch \cite{Spradlin}. But while de Sitter may be a reasonable model for our Universe in the asymptotic future, it cannot be a good fit for matter and radiation dominated eras nor the Big Bang itself.}

An expanding Big Bang Universe invalidates the Perfect Cosmological Principle, but one could still ask the question: what is the {\em most} Copernican paradigm for cosmology that is compatible with an expanding Universe? The answer to this question is provided by the Friedmann-Lema\^{i}tre-Robertson-Walker (FLRW) metric \cite{WeinbergOldBook, Ellis-Maartens-McCallum--Book}. The FLRW ansatz allows for time dependence, but assumes that the universe is maximally symmetric (ie., homogeneous and isotropic) on spatial slices. This is the Cosmological Principle. FLRW ansatz is the starting point for modern cosmology, and the standard model of modern cosmology (the flat \lcdm\  concordance model) is a specific instantiation of it. 

\lcdm\  cosmology has been a huge success story overall in providing a concordant model for the Universe across datasets at different redshifts. But in the last two decades or so, multiple worries have emerged in the cosmological standard model. Most striking among them is the sustained tension that has only gotten worse over the years, in the measured value of the local Hubble constant $H_0$  \cite{crisis, Intro0,Eleonora-et-al-review-1, SNOWMASS-2022}. 
The inferred value of $H_0$ from early cosmology  (e.g. as reported by Planck mission, $\sim 67.4\pm 0.5$ km/s/Mpc \cite{Planck-2018}) differs from those arising from local measurements ($\sim 72-73$ km/s/Mpc see e.g. \cite{SNOWMASS-2022}). This discrepancy is often quoted to be at the $4$-$ 6\ \sigma$ \cite{SNOWMASS-2022,Eleonora-combined-analysis} level. $H_0$ tension is one among many tensions challenging the concordance between early and late Universe datasets (see \cite{SNOWMASS-2022} and references therein), giving rise to what has been called a crisis in $\Lambda$CDM cosmology \cite{crisis}.  Given this state of affairs, it is interesting to explore how one should modify models of cosmology if one were to take current data at face value.  While it is not yet a certainty that the systematics are under control and that the data is beyond question, it does motivate us to evaluate what might be the weakest points in our current models.  In particular, our own recent work motivated by $H_0$ tension has increasingly lead us to consider the possibility that the FLRW paradigm may be breaking down \cite{Anjan, Beyond-FLRW-review}.

Continuing in this vein, in this paper we will identify the most symmetric paradigm that generalizes FLRW cosmology, in which a cosmic flow can be incorporated.
The motivation is that during the last decade or so, tentative evidence has accumulated (see eg. \cite{SubirEtc, Rameez, SNe-hemisphere-anomaly, Subir2, QSO-hemisphere-anomaly}) suggesting that the dipole in the CMB may have a non-kinematical component  \cite{Ellis-Baldwin}.\footnote{Let us emphasize that ``non-kinematic" does not necessarily mean ``intrinsic" CMB dipole. Ellis-Baldwin\cite{Ellis-Baldwin} simply identify the possibility that if the dipole is due to our local random motions then it will affect all distant sources (CMB as well as say, quasars) in a simple way. But if this is not found to be the case observationally, it does not automatically mean that the CMB dipole is intrinsic. It can (for example) be the case that the cosmic fluid has red-shift dependent flows at cosmological scales. This possibility is what we are exploring in this paper. In particular, note the papers \cite{Saha, Quartin} which argue that the CMB dipole cannot be intrinsic, by considering a special relativistic boost between the CMB and us. This does not affect our discussions, because all this implies is that the dipole is due to a relative flow/motion between the CMB and us -- there is no statement about what happens to the flow at intermediate redshifts between now and last scattering. The existence of such a flow will affect the sources at intermediate redshifts and can in principle lead to a non-trivial Ellis-Baldwin test result.}  We wish to write down the most Copernican paradigm that can incorporate this. We will write down the explicit metric, stress tensor and equations of motion for these spacetimes and study these cosmologies for some interesting classes of equations of state. We will find that the flows in these Universes do {\em not} have to die down even when there is late-time acceleration. 

\paragraph{What we do in this paper and our main result.} 
The search for the most symmetric Universe that allows  cosmic expansion and accommodates a flow, will inexorably lead us to a specific (sub-)class of the so-called tilted homogeneous cosmologies \cite{King}. The cosmologies we find are spatially homogeneous but anisotropic tilted Bianchi models, that fall specifically within Bianchi V or VII$_h$ classes and have an axial $U(1)$ isotropy around the flow. We will explain the various technical words in the above sentence as we go along. Useful general discussions and references on anisotropic homogeneous models can be found in the lectures \cite{Ellis-lectures} as well as in chapter 18 of \cite{Ellis-Maartens-McCallum--Book}. The class of models we will arrive at are specified by an overall scale factor, a shear as well as a tilt (on top of the pressure and density of the fluid). Shear captures the anisotropy in the background cosmology and tilt is essentially the flow velocity of the cosmic fluid. Since cosmic flows can result in dipoles in the observed distributions of sources in the sky, we call these ``dipole cosmologies". We work out field equations for these cosmologies and find that there are four ordinary differential equations (ODEs); two of them are extensions of Friedmann equations and the other two govern the shear and the tilt. These equations form a closed system once one adds the (effective) Equation of State (EoS) of the cosmic fluid. 

The fact that the equations of motion are ODEs\footnote{This is a consequence of the homogeneity on spatial slices in dipole cosmology, that follows from our Copernican prior.} means that the system is essentially as tractable as FLRW. We analyze these equations and plot the time evolution of the physical quantities for various illustrative choices of the EoS. A key technical idea we invoke in order to get some general understanding of these cosmologies is to work with a \textit{total} effective EoS that is a function of time, instead of fluid mixtures as is often done in FLRW. Our main, probably unexpected, result is that the dipole flow in expanding dipole cosmologies can increase, even in cases where the shear (anisotropy in the metric) dies off at late times. What is particularly striking is that this is quite generically true (in a sense that we will elaborate) even in Universes which exhibit late time acceleration. In other words, even as the metric sector is isotropizing as expected from the cosmic no-hair theorem \cite{Wald-cosmic-no-hair}, the flows in the matter sector in an accelerating Universe can lead to non-negligible cosmic dipoles. Note that our claim here is {\em not} that only accelerating dipole cosmologies are phenomenologically interesting -- that is a question that requires a re-evaluation of all data in the dipole cosmology setting, and is beyond the scope of this paper. Our observation is that {\em even } in accelerating Universes, dipole flows cannot be ignored -- they are not washed away by cosmic no hair theorems and the like.

\paragraph{Organization of the paper.} In section \ref{sec:dipole-cosmological-principle}, we discuss how Copernican principle should be amended to accommodate a bulk flow, we call it the ``Dipole Cosmological Principle''.  We are led to a specific tilted Bianchi model as a result. In section \ref{sec:3}, we write down the explicit form of the metric and the (tilted) energy-momentum tensor of dipole cosmology.  Here we also work out the basic dynamical equations, which are extensions of Friedmann equations to the case of dipole cosmology. In section \ref{sec:4-numerical-analysis}, we analytically and numerically study various specific models (ie., choices of EoS) and set the stage for model building within the dipole cosmology paradigm. We find that a useful tool for extracting general lessons is to treat the total EoS as a chosen function of time. Analysis of this section reveals that the tilt need not die off, even though the shear can go to zero as the Universe expands. In particular, we find multiple examples where the total EoS tends to $w_{tot}=-1$ at late times and the Universe accelerates, concurrently with an increasing tilt. In section \ref{sec:5}, we study more closely the tilt-growth phenomenon. Through a perturbative late time expansion  around a Universe with a $w(t) \rightarrow -1$ equation of state, we first show that tilt-growth is a generic feature in a late-time accelerating dipole cosmology for a range of initial values of the parameters. Next, we briefly discuss scenarios where tilt growth can happen not only at late times, but at intermediate epochs. We also identify examples where tilt can have a maximum as it evolves in time.  In section \ref{sec:discussion}, we close by making some concluding remarks, where we briefly discuss cosmological implications of our results. 


\section{Copernican Priors: Dipole Cosmological Principle}\label{sec:dipole-cosmological-principle}

How does one go about constructing a cosmology with a cosmic flow?  This is where we will invoke a lesson from history and adopt a Copernican approach. As we already discussed, the Copernican principle as applied to cosmology is a principle of ignorance. {\em It is a working assumption that dictates that the spacetime we work with is the most symmetric one, compatible with our priors.} When we did not know that the Universe was expanding, the Perfect Cosmological Principle  was such a symmetry principle about the metric. After the discovery of cosmological expansion, this was no longer a tenable prior, and the (ordinary) Cosmological Principle became the new symmetry principle. This immediately implied the familiar FLRW metric as the unique ansatz for our Universe, and from there we were inexorably led to the Friedmann equations. 

If cosmic flow is a real thing, the natural path for us to take then, is to look for a Copernican cosmology where the flow is a part of the prior. In other words, we will seek a metric that has the maximum amount of symmetry compatible with both cosmic expansion, as well as the existence of a flow. We will call this the {\em Dipole Cosmological Principle}.  

Maximal symmetry on a spatial 3-slice implies that there is a group of symmetries of dimension 6 acting transitively on the spatial 3-slice. The number of generators is 6 because the maximal number of Killing vectors in 3 dimensions is 6 -- we will use the words symmetry and isometry interchangeably. Now, {\em transitive} action means that one can get from any point on the 3-slice to any other point via the action of some (possibly non-unique) element of the group -- the entire 3-slice can be viewed as being in the {\em orbit} of the group. It should be intuitively clear that the minimum dimensionality of a group so that transitive action can be realized on a 3-slice, is 3. A 3-slice on which a group has transitive action is called a {\em homogeneous} 3-slice. This is the natural definition of a homogeneous (sub-)space because the ability to go from any point to any other point by the action of a symmetry is the intuitive notion of homogeneity. 

At each point on the 3-slice we can also define the {\em isotropy} group, as the subgroup of the isometry group that leaves the point fixed. An example of an isotropy that is useful to keep in mind, is a rotation (when it is an isometry). Clearly, rotation around a point does not move the point. The sum of dimensionalities of the surface of transitivity and the isotropy group at each point, is equal to the dimensionality of the isometry group. This is again intuitive; moving along the surface is accomplished precisely by those generators of the group that do not leave the point fixed. In any event, for a maximally symmetric 3-slice therefore, the isotropy group is 3-dimensional. This is simply the full rotation group in 3-dimensions, $SO(3)$, which has 3 generators. With this $SO(3)$ isotropy group on a homogeneous 3-slice, we are led uniquely to the FLRW metric \cite{WeinbergOldBook, Ellis-Maartens-McCallum--Book}. 

Maximally symmetric 3-slices do not allow flows \cite{King}. So we need to make our 3-slices less symmetric. Our Copernican philosophy suggests that the thing to do here is to consider a homogeneous space (ie., a 3-slice where a symmetry with at least 3-generators is  acting transitively) which allows the maximum amount of symmetry compatible with the existence of a flow. It is intuitively plausible that if there is a flow and therefore a preferred spatial direction, the isotropy group can at most be $U(1)$ -- the group of axial rotations around the flow direction. Indeed, this turns out to be the case, because there are no 2-dimensional subgroups of $SO(3)$. The full isometry group, together with the simply transitive part, is therefore  of dimension 4. Such spacetimes, which allow a spatially homogeneous 3-slice foliation with an axial $U(1)$ isotropy around each point, are the subject of this paper. We will call them dipole cosmologies because the flow which is taken along the $U(1)$ rotation axis, will find the natural interpretation as a non-kinematic contribution to the cosmic dipole.

Together with the above assumptions about the isometries of spacetime, we will also take that the stress tensor is of a perfect fluid form. This is again in parallel to the FLRW case. But there is a  crucial difference when we want to include a flow that can mimic a non-kinematic dipole. When there is a flow, the fluid velocity 4-vector $u^a$ is not orthogonal to the 3-slices of homogeneity. This is sometimes referred to as a ``tilt'' \cite{King, Ellis-Maartens-McCallum--Book}. We will use this word synonymously with (cosmic) flow.  

To summarize, the Universes we will be interested in, are those that are homogeneous on 3-slices, have an axial isotropy group $U(1)$ at each point, and have a fluid flow 4-vector that is {\em not} orthogonal to the surfaces of homogeneity. This is what we will call, {\em Dipole Cosmology}. It turns out that metrics in this class can be viewed as belonging to Type V or Type VII$_h$ Bianchi classes -- the two are indistinguishable when there is an axial $U(1)$. We will write down the most general metric, stress tensor and equations of motion that are compatible with these assumptions in the next section. The resulting equations are to dipole cosmology, what the Friedmann equations are to FLRW cosmology.

Spacetimes with an axial $U(1)$ isotropy, but without the assumption of spatial homogeneity on the 3-slice, were considered in \cite{Stewart-Ellis}. What we call axial isotropy was called ``local rotational symmetry'' (LRS) there. We will only be interested in 3-slices that are homogeneous, and therefore our spacetimes are special cases of those discussed in \cite{Stewart-Ellis}. In \cite{King} spatially homogeneous 3-slices without axial isotropy were considered, with a brief discussion of the axially isotropic special case. Our goal will be to elaborate on this special class, write down explicit metrics, stress tensors and equations of motion and discuss the preliminary phenomenology of such cosmologies. Spatially homogeneous cosmologies with reduced isotropy groups were also considered in \cite{MacCallum} but the possibility of flows were excluded because the fluid velocity was assumed to be orthogonal to the 3-slices of homogeneity.

\section{Dipole Cosmology}\label{sec:3}

Homogeneous Universes with reduced isotropy have been studied in one form or another for a long time  \cite{Ellis-lectures, Ellis-Maartens-McCallum--Book}, see \cite{Tsagas:2021tqa} and references therein for more recent work. The list of 3-dimensional homogeneity algebras was made 120 years ago by Bianchi, purely as a mathematical problem. So these cosmologies are called Bianchi cosmologies. Our primary resources will be the papers by Ellis and collaborators \cite{MacCallum, King, Stewart-Ellis, Ellis-lectures} which are reviewed and discussed in \cite{Ellis-Maartens-McCallum--Book}. The formalism they work with is somewhat baroque, and uses a local frame language to connect with the Bianchi symmetry classes. We wish instead to work with a metric language that is accessible and intuitive, in particular we want a set of equations that generalize the Friedmann equations. It is possible to systematically translate the results of \cite{MacCallum, King} into a metric language, and we will describe this in detail in a companion paper \cite{KM}. There we will also discuss more general dipole cosmologies than the axially isotropic (Copernican) one we consider in this paper. But for the goals of the present paper, where we wish to retain as much symmetry as we can, we will simply present the final results for the metric, stress tensor and equations of motion. This is the goal of this section. It is possible to adapt the frame formalism of \cite{MacCallum, King, Ellis-lectures} to obtain the results we present below; we will do this in a somewhat more general setting in \cite{KM}. 

\subsection{Background Metric and Tilted Cosmic Fluid} 
The most general metric compatible with the Copernican assumptions of the previous section (ie., the Dipole Cosmological Principle) is
\bea\label{DipoleMetric}
    ds^{2} = -dt^{2} + X^{2}(t)\ dz^{2} + Y^{2}(t)\ \exp(-2A_{0}\ z)\big(dx^{2}+dy^{2}\big) 
\eea
The above metric makes the 4 Killing vectors on a constant $t$ slice explicit: $\partial_x, \partial_y, x\partial_y-y\partial_x$ and $K=A_0 (x\partial_x+ y\partial_y)+\partial_z$. Here, $A_0$ is a constant. For $A_0=0$ it reduces to a Binachi I model which does not allow a tilt. Since we will be interested exclusively in $A_0\neq 0$, it is worth noting that the parameter $A_0$ which has the dimension of inverse length,  may be set to one upon a scaling of the $z$ coordinate and a redefinition of $X(t)$. We will, however, often keep $A_0$ explicit as it helps with tracking the dimensions of various terms appearing in the equations. 

The metric \eqref{DipoleMetric} falls into Bianchi V and VII$_h$ classes, see eg. chapter 18 of \cite{Ellis-Maartens-McCallum--Book}. A constant time slice of the above  describes (a patch of)  a 3-dimensional hyperboloid $H^3$ with $SO(3,1)$ isometry. In other words, for the special cases where $X(t)/Y(t)=$ const., the above metric describes an open FLRW Universe with $k=-1$. Put differently, the line element \eqref{DipoleMetric} may be viewed as the deformation of an open FLRW spacetime with a nonzero cosmic shear $\sigma:=\dot{X}/X-\dot{Y}/{Y}$. 
In the coordinates adapted here (for the $X(t)/Y(t)=$ const. case), we have an $R^2$ slicing of this $H^3$. Explicitly the metric on a constant $t_0$ slice is
\begin{equation}
\begin{split}
ds^2=\frac{X(t_0)^2}{A_0^2\tilde{z}^2}\left[d\tilde{z}^2+d\tilde{x}^2+d\tilde{y}^2\right]\nonumber
\end{split}
\end{equation}
where $\tilde{z}= \frac{e^{A_0z}}{A_0}$ and $\tilde{x}=\frac{Y(t_0)}{X(t_0)} x, \tilde{y}=\frac{Y(t_0)}{X(t_0)} y$. This is the metric for an $H^3$ in the Poincare patch. The $R^2$ slicing is particularly suitable for our case, as it singles out one spatial direction ($z$ direction) along  the cosmic flow.

 In $(t,z, x,y)$ coordinates that are adapted to the form of  metric \eqref{DipoleMetric}, the perfect fluid stress tensor with energy density $\rho$ and pressure $p$ with a flow (aka ``tilt'') takes the form \cite{KM}
\begin{equation}
\small{
    T_{ab} = \left(
\begin{array}{cccc}
 {\rho +(\rho+p) \sinh ^2\beta} & -(\rho +p) X \sinh \beta \cosh \beta & 0 & 0 \\
 -(\rho +p) X\sinh \beta \cosh \beta & \left(p+(\rho +p) \sinh ^2\beta \right) X^2 & 0 & 0 \\
 0 & 0 & p\ e^{-2 A_{0} {z}}\ Y^2 & 0 \\
 0 & 0 & 0 & p\ e^{-2  A_{0} {z}}\ Y^2 \\
\end{array}
\right)}
\end{equation}
or equivalently,
\begin{equation}\label{tilted-EM-up-down}
T^{a}{}_{b} 
= \text{diag}(-\rho, p,p,p)+(\rho+p)\sinh\beta\left(\begin{array}{cccc}
- \sinh \beta &  X  \cosh \beta & 0 & 0 \\
-X^{-1} \cosh \beta &  \sinh\beta  & 0 & 0 \\
 0 & 0 & 0 & 0 \\
 0 & 0 & 0 & 0\\
\end{array}
\right)
\end{equation}
Note that even though the underlying fluid is a perfect fluid, this stress tensor has ``imperfect" terms due to the flow, {manifested in the parameter $\beta (t)$ and that $T^a{}_b$ has off-diagonal elements. Here the independent functions are $X, Y, \rho, p$ and $\beta$, and they are all functions of only $t$. We have suppressed the dependence on $t$ in $p, \rho$ and $\beta$ in the expressions above for aesthetic reasons. Compared to the usual FLRW set up,  we have two extra functions of $t$ as degrees of freedom. These can be viewed as the extra anisotropic scale factor $X(t)$ which is responsible for breaking the isotropy from $SO(3)$ to $U(1)$, and the fluid dipole flow velocity or tilt, $\beta(t)$.

It is worth noting that for the $p=-\rho$ case $\beta$ drops out of the stress tensor and we get a usual perfect fluid, ie., cosmological constant is compatible with any flow or tilt. We will consider  equations of state $w(t):=p/\rho$ with a chosen time-dependence later in this paper. In particular, we also study  $w(t)\rightarrow -1$  as a late-time limiting value of physical interest. The connection between such a scenario, and the addition of a genuine cosmological constant will be clarified eventually.  We will see later that flow velocities do not have to die down with time, even though in the Universes can be accelerating. Note also that trace of $T_{ab}$ is $\beta$ independent, $T^a{}_a=-\rho+3p$ and hence radiation remains trace-free}.


\subsection{Field Equations} 

The equations of motion from the metric and stress tensor above are ordinary differential equations of $t$, just as for FLRW, but we expect to get two more equations on top of the usual Friedmann equations because of the two new functions. The explicit evaluation of the Einstein equations confirms this. We consider Einstein equations with an explicit cosmological constant on top of the tilted perfect fluid, 
\bea
G_{ab}+\Lambda \ g_{ab}= T_{ab}, \label{cosmoC}
\eea
and find two second order equations
\begin{subequations}
\begin{align}
    \frac{\ddot{X}}{X} + 2\frac{\dot{X}}{X}\frac{\dot{Y}}{Y} - 2\frac{A_{0}^{2}}{X^{2}} = &\frac{1}{2}(\rho-p) + (\rho + p)\sinh^{2}{\beta} + \Lambda \label{FW3}\\
    \frac{\ddot{Y}}{Y} + \big(\frac{\dot{Y}}{Y}\big)^{2} + \frac{\dot{X}}{X}\frac{\dot{Y}}{Y} - 2\frac{A_{0}^{2}}{X^{2}} =& \frac{1}{2}(\rho-p) + \Lambda \label{FW4}
\end{align}
\end{subequations}
and two first order equations
\begin{subequations}
\begin{align}
    \frac{2A_{0}}{X}\big(\frac{\dot{X}}{X}-\frac{\dot{Y}}{Y}\big)=& (\rho + p)\sinh{\beta}\cosh{\beta} \label{FW1}\\
    2\frac{\dot{X}}{X}\frac{\dot{Y}}{Y} + \big(\frac{\dot{Y}}{Y}\big)^{2} - \frac{3 A_{0}^{2}}{X^{2}} =& \rho +(\rho+p)\sinh^{2}{\beta} + \Lambda.\label{FW2}
\end{align}
\end{subequations}
These four equations are our generalizations for the Friedmann equations, which as may be recalled, contain two equations (one second order equation and one first integral). The two extra equations are necessary to incorporate the two new extra functions we mentioned above. 

As in the case of the usual Friedmann equations, here also it is possible to replace the second order equations with the conservation law for the stress tensor.
We will indeed find it convenient to work with them in our (numerical) evolution for the various equations of state that we will consider. The independent equations obtained from the covariant conservation of the stress tensor are:
\begin{subequations}
\begin{align}
\dot{\rho}+(\rho+p)\Big(\frac{d}{dt} \log ({X}{Y}^2 \cosh \beta)-\frac{2 A_0}{X}\tanh \beta\Big)=&0 \label{Con1} \\
\dot{p}+(\rho+p)\frac{d}{dt} \log ({X}\sinh \beta)=&0 \label{Con2}
\end{align}
\end{subequations}
As in the case of FLRW, it is possible to explicitly check that one can reproduce these equations by taking one more derivative of the first derivative constraints (here \eqref{FW1} and \eqref{FW2}) and then eliminating the second derivatives using the second order equations (here \eqref{FW3} and \eqref{FW4}). Interestingly enough, we found that this process is algebraically more tricky (at least, more tricky than we expected) in the case of the Dipole Cosmology system, than the usual Friedmann system. 

It is sometimes useful to write the above equations in a slightly different notation. To this end, instead of $X(t), Y(t)$ we introduce, 
\begin{equation}\label{XY-ab}
X:=a(t)\ e^{2b(t)}, \qquad Y:=a(t)\ e^{-b(t)}  
\end{equation} 
where $a(t)$ is the overall scale factor ($a^3=XY^2)$ and $b(t)$ parameterizes the anisotropy ($e^{3b}=X/Y$). One may then define Hubble expansion rate $H(t)$ and the shear $\sigma(t)$ as usual
\begin{equation}\label{H-sigma}
H:=\frac{\dot{a}}{a},\qquad \sigma:=3\dot{b} 
\end{equation}
The field and continuity equations then take the form
\begin{subequations}\label{EoM-H-sigma}
\begin{align}
\dot{H}+3H^2-2\frac{A_0^2}{a^2} e^{-4b}&=\frac12(\rho-p)+\frac13 (\rho+p)\sinh^2\beta+\Lambda\label{EoM-H-sigma-a} \\
\dot{\sigma}+3H\sigma&= (\rho+p)\sinh^2\beta \label{shear-EoM}\\
H^2-\frac19\sigma^2-\frac{A_0^2}{a^2} e^{-4b}&=\frac{\rho}{3}+\frac13 (\rho+p)\sinh^2\beta+\frac{\Lambda}{3} \label{EoM-H-sigma-c}\\
\frac{2A_0}{a}e^{-2b} \sigma &= (\rho+p)\sinh\beta\cosh\beta \label{EoM-H-sigma-d}
\end{align}
\end{subequations}
\begin{subequations}\label{EoM-continuity}
\begin{align}
\dot{\rho}+3H(\rho+p)=&-(\rho+p)\tanh\beta(\dot{\beta}-\frac{2A_0}{a} e^{-2b}) \label{Con1-1} \\
\dot{p}+H(\rho+p)=& -(\rho+p)\left( \frac23\sigma+\dot{\beta}\coth{\beta}\right). \label{Con2-1}
\end{align}
\end{subequations}

Before performing a more detailed analysis of the above equations, some comments are in order. 
\begin{itemize}
\item As \eqref{FW1} or \eqref{shear-EoM} \& \eqref{EoM-H-sigma-d} manifest, for $\beta=0$ we get either $A_0=0$ or $X(t)/Y(t)=$ const. (or $\sigma=0$). In our analysis we do not consider $A_0=0$ case, which yields a Bianchi I model that does not allow for a tilted (dipole) cosmology. For $\sigma=0$ case the equations reduce to those of a usual open FLRW universe with untilted matter. 
\item The other special case is when $\rho+p=0$. In this case, the tilt $\beta$ drops out of the equations and hence the value of $\beta$ remains unspecified by the equations. For this case, \eqref{shear-EoM} \& \eqref{EoM-H-sigma-d} imply $\sigma=0$. The solution is a (A)dS$_4$ space in an $H^3$ slicing, and the $H^3$ itself is written in an $R^2$ slicing, as discussed above.
\end{itemize}

\section{Phenomenology of Schematic Models}\label{sec:4-numerical-analysis}

The equations above are direct generalizations of the Friedmann equations, and therefore it is interesting to see whether they allow universes that are broadly similar to ours. Here, we must confront a problem that is implicit in the FLRW paradigm as well -- namely that the system of differential equations is incomplete. In the Friedmann system, we have two independent ODEs, but three unknowns: the scale factor $a(t)$, the pressure $p(t)$ and the density $\rho(t)$. In our case, we have four independent ODEs, but five unknowns $X(t), Y(t)$ (or $a(t), b(t)$) and $p(t) ,\rho(t), \beta(t)$. 

The way in which this problem is usually addressed in the FLRW setting is by postulating the idea of an ``equation of state" (EoS). We assume that we have extra information about $w := p/\rho$. In $\Lambda$CDM model (for example), the EoS information goes in via the following sequence of assumptions:
\begin{itemize}
\item The total energy-momentum tensor appearing in the right-hand-side of Einstein equations is  the sum of separate perfect fluid stress tensors for different components, ie., for an $N$ component cosmic fluid, $T_{ab}= \sum_{i=1}^N T^i_{ab}$. 
\item Each of these component stress tensors are individually covariantly conserved, ie. $\nabla^a T^i_{ab}=0$ for any $i$. These two assumptions mean that different components do not interact with each other, or more precisely, they only interact gravitationally. 
\item The component EoS $w_i$ are typically assumed to be constants.  For example,  $w= 1, 1/3, 0, -1/3$ and $-1$ respectively correspond to stiff fluid, radiation, (dark and baryonic) dust, spatial curvature and vacuum energy. 
\end{itemize}
Together with these assumptions, the Friedmann system now becomes solvable. We will discuss suitable generalizations of this to the tilted setting in what follows, but we start with the simplest dipole cosmology example of a single fluid with a constant EoS. For simplicity we will also set the cosmological constant $\Lambda=0$ in the next subsection, but we will discuss  more general cases eventually.

\subsection{Models with Constant EoS}\label{sec:dipole-const-w-Eos}

As the first example we consider dipole cosmologies with $w$=const. where $-1<w\leq 1$, with no additional cosmological constant. The special cases of particular interest are $w=0$ describing a pressureless matter (dust) and $w=1/3$ describing radiation. 

\begin{figure}[H]
    \centering
    \subfloat[\label{fig:aConstant}]{{\includegraphics[width=8.0 cm]{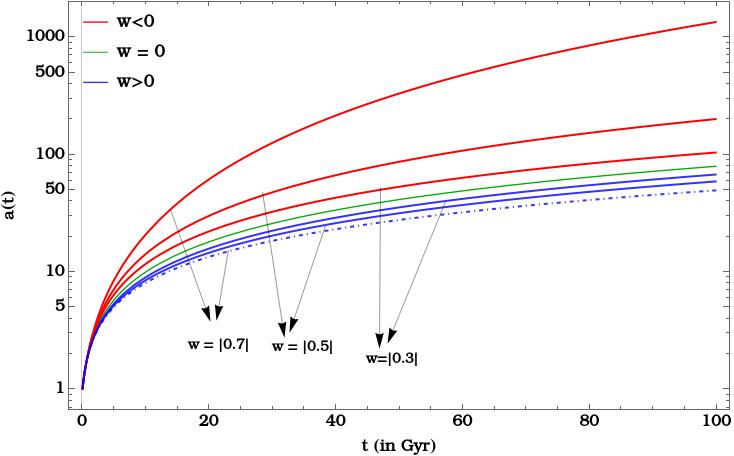} }}\hfill
    \subfloat[\label{fig:rhoConstant}]{{\includegraphics[width=8.0 cm]{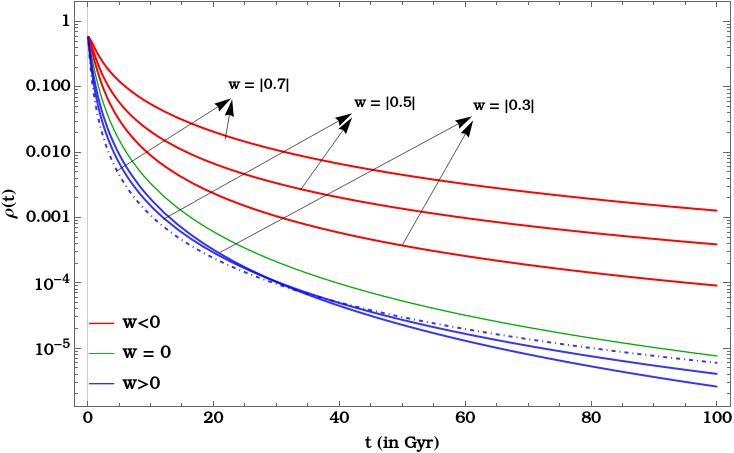} }}
  \\
\centering
    \subfloat[\label{fig:betaConstant}]{{\includegraphics[width=8.0 cm]{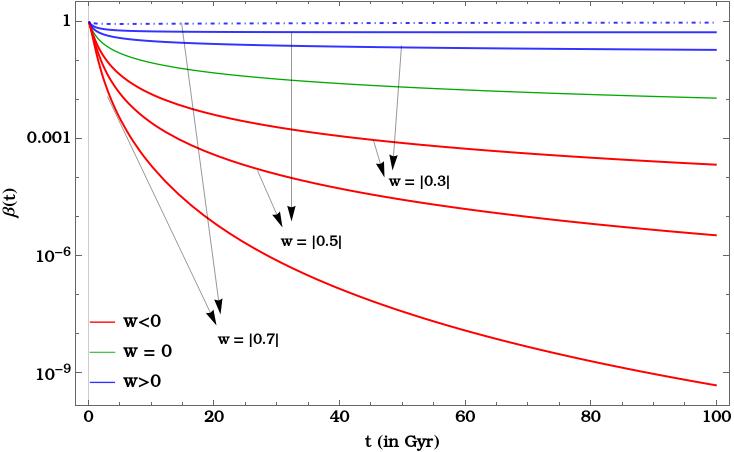} }}\hfill
    \subfloat[\label{fig:sigmaConstant}]{{\includegraphics[width=8.0 cm]{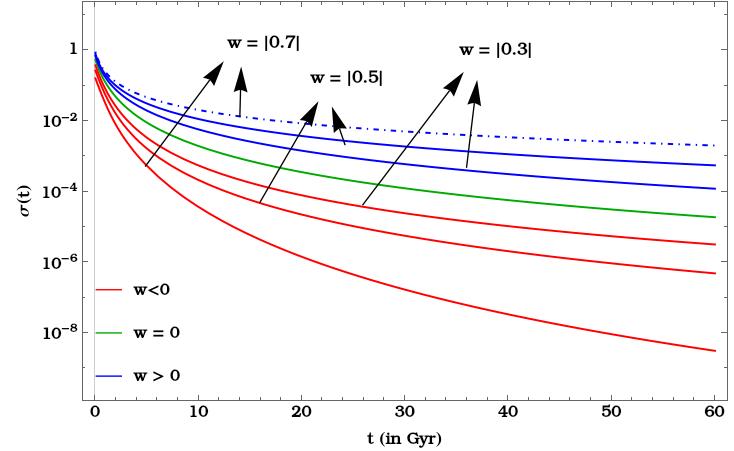} }}
    \caption{Evolution of overall scale factor, tilt, density and shear for constant equations of state within $-0.7 \leq w \leq 0.7$ for initial values $X_{in} = Y_{in} = 1$, $\rho_{in} = 0.6$, $\beta_{in} = 1$. We set our initial conditions at $t = 0.01$ Gyr. Because of the crowding of curves for larger $w$, some of the details are split off into Figure \ref{fig:ConstEoSbeta}.}
    \label{fig:ConstantEOS}
\end{figure}

\paragraph{$w=0$, dipole dust model.}  For this case \eqref{Con2-1} and \eqref{Con1-1} imply,
\bea \label{dustconstraint}
  X\sinh{\beta}=const., \qquad \dot{\beta}\coth\beta=-H-\frac23\sigma
\eea
This shows that for an expanding universe with pressureless dust the tilt angle $\beta$ always decreases with time. Moreover, one also finds that
$\sigma= C \rho \cosh\beta$, where $C$ is an integration constant. The tilt $\beta$ and the shear $\sigma$ drop to very small values at late times and the model essentially reduces to an FLRW dust solution, in which $\rho\sim a^{-3}\sim t^{-2}$. These features are visible in $w=0$ plots in Figure \ref{fig:ConstantEOS}.

\paragraph{$w=1/3$, dipole radiation model.} As the next example we consider radiation $p=\rho/3$. For this case \eqref{Con2-1}  and \eqref{Con1-1} imply
\begin{equation}
\dot{\beta}(3\coth\beta-\tanh\beta)=-\frac{2A_0\tanh\beta}{X}-2\sigma,\qquad    \rho X^4 \sinh^4\beta= const.
\end{equation}
The RHS of the first equation is always negative while the coefficient of $\dot{\beta}$ is always positive. Therefore, $\dot\beta<0$, so the tilt is always decreasing. For late times $\rho a^4\sim$ const and $a\sim t^{1/2}$. The numerically evolved system has been plotted in Figure \ref{fig:ConstantEOS}.

\paragraph{Generic constant $w$ case.} When $p=w \rho$, \eqref{Con2-1} and \eqref{Con1-1} imply,
\begin{subequations}\label{const-w-dipole-1}
\begin{align}
&\dot{\beta}\big(\coth\beta-w \tanh \beta\big)=(3w-1)H-\frac23\sigma-\frac{2w A_0\tanh\beta}{X}, \label{beta-growth-w-const}
\\
&\rho^{\frac{w}{1+w}} X \sinh\beta= C= const. \label{const-w-dipole-rho-X-beta}
\end{align}
\end{subequations}
Since $-1<w\leq 1$, and recalling that $\coth\beta>1, \tanh\beta <1$ once we assume $\beta>0$, the coefficient of $\dot\beta$ term is always positive. The $\tanh\beta/X$ term can be positive or negative depending on the sign of $w$. Nevertheless, this term becomes insignificant at late times due to the expansion ($X$ becoming very large). For $w\leq 1/3$, the first term is also negative and hence $\dot{\beta}<0$ at late times. For $w>1/3$ the first term of the RHS of \eqref{beta-growth-w-const} is positive and depending on the other details of the evolution, there is a possibility that $\dot{\beta}>0$. 

We note that the cosmic acceleration for this case is given by 
\begin{equation}\label{cosmic-acceleration-const-w}
\frac{\ddot{a}}{a}=   \dot{H}+H^2=-\frac{\rho}{6}(1+3w)-\frac29 \sigma^2-\frac{\rho}{3}(1+w)\sinh^2\beta.
\end{equation}
As we see, the first term in the RHS is the usual FLRW term, which yields an accelerating Universe for $w<-1/3$. The last two terms, and in particular the tilt term, are negative definite and hence to get acceleration the first term should overcome the last two negative terms. However, our numerical analysis (cf. Figure \ref{fig:ConstantEOS}) shows that for $w<-1/3$ the last two terms die down quite fast at late times, where the FLRW dynamics dominates. As already discussed, for decelerating cosmologies with $w>1/3$, equation  \eqref{const-w-dipole-1} suggests there is the possibility of tilt growth. Our results are compatible with the earlier analysis presented in \cite{Coley-2006}.
In our numerical analysis we have been able to get tilt growth for $w$ as low as $\sim 0.4$ by tuning the initial conditions. Whether one can get tilt growth for lower $w$ (all the way down to $w=1/3$) by some other appropriate initial conditions remains to be explored in future works. Even though we will not present the details, let us also mention that adding a positive $\Lambda$ to the system does not change these observations qualitatively. We still find that $\beta$-increase may occur when $w > 1/3$.  In fact, $\Lambda$ makes the increase in $\beta$ more robust numerically --  by tuning initial conditions we were able to identify increasing $\beta$ at $w$ as low as $\sim 0.35$.

For completeness we also quote  the full set of field equations for a dipole cosmology with a single cosmic fluid of constant $w$. They take an interesting and (perhaps surprisingly) simple form: 
\begin{subequations}
\begin{align}
&     \sigma = \frac{C (1+w)}{2A_0} \rho^{\frac{1}{1+w}} \cosh\beta, \label{sigma-const-w}\\
&     H=\frac{A_0}{C}\rho^{\frac{w}{1+w}}\left[1+\frac{C^2w}{3A^2_0}\rho^{\frac{1-w}{1+w}}\right]^{1/2}\ \sinh\zeta, \label{H-zeta}\\
&    \left(1+\frac{C^2(1+w)}{6A_0^2}\rho^{\frac{1-w}{1+w}}\right)\cosh\beta= \left[1+\frac{C^2w}{3A^2_0}\rho^{\frac{1-w}{1+w}}\right]^{1/2}\ \cosh\zeta, \label{rho-zeta}\\
 &    \frac{d}{dt}\ln\left(\rho^{\frac{1}{w+1}} a^3\cosh\beta\right)=\frac{2A_0}{C} \rho^{\frac{w}{w+1}}\tanh\beta\sinh\beta \label{H-beta-w}
\end{align}\end{subequations}
where  $\zeta$ is an unknown function to be determined upon plugging \eqref{H-zeta}, \eqref{rho-zeta} into \eqref{H-beta-w}. Then, \eqref{sigma-const-w} may be used to find $\sigma$ or $b$.

\begin{figure}[H]
    \centering
    \subfloat[\label{fig:w1}]{{\includegraphics[width=8.0 cm]{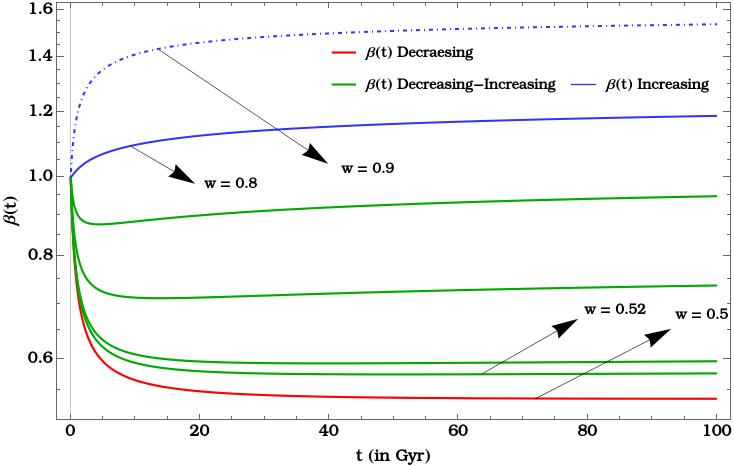} }}\hfill 
    \subfloat[\label{fig:aVar1}]{{\includegraphics[width=8.0 cm]{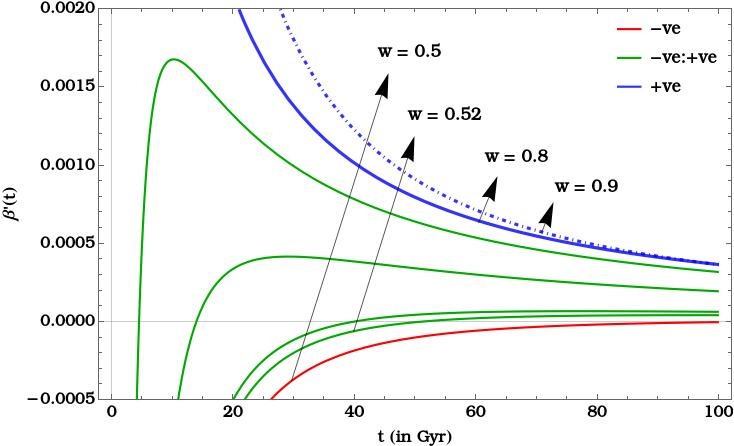} }}\\ \subfloat[\label{fig:w1h}]{{\includegraphics[width=8.0 cm]{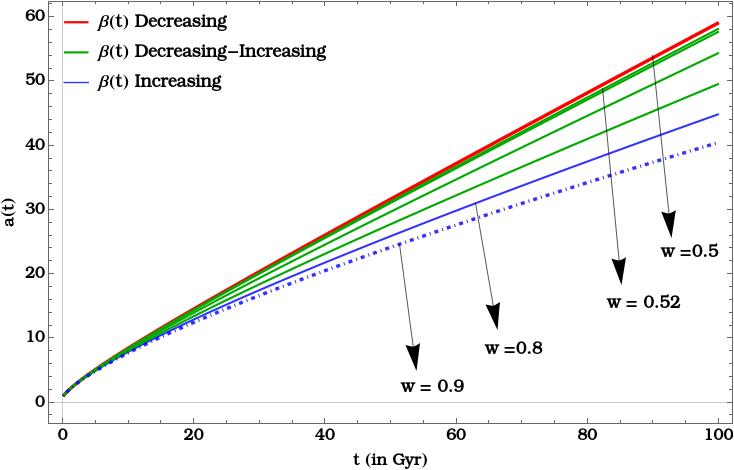} }}\hfill
     \subfloat[\label{fig:rhoVar1}]{{\includegraphics[width=8.0 cm]{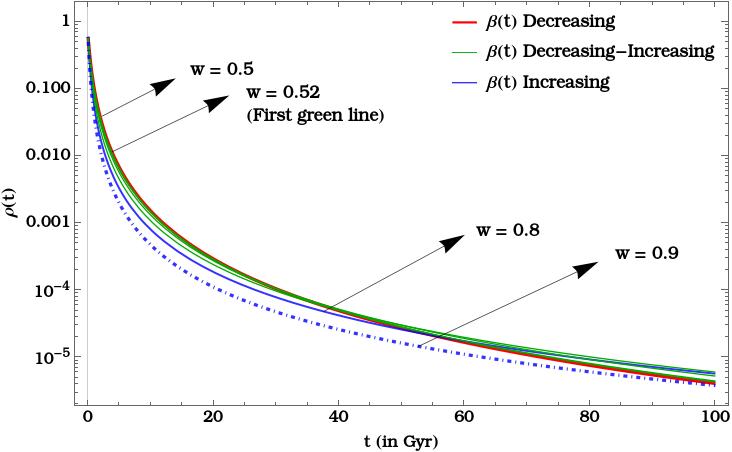} }}\\ 
      \subfloat[\label{fig:betVar1}]{{\includegraphics[width=8.0 cm]{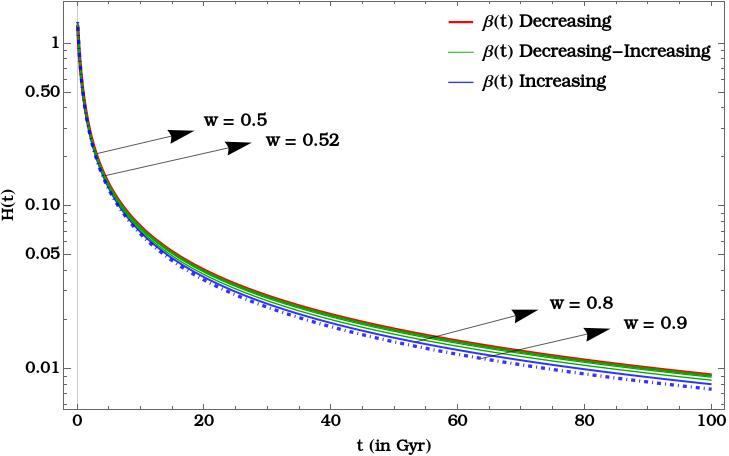} }}\hfill
      \subfloat[\label{fig:sigmaVar1}]{{\includegraphics[width=8.0 cm]{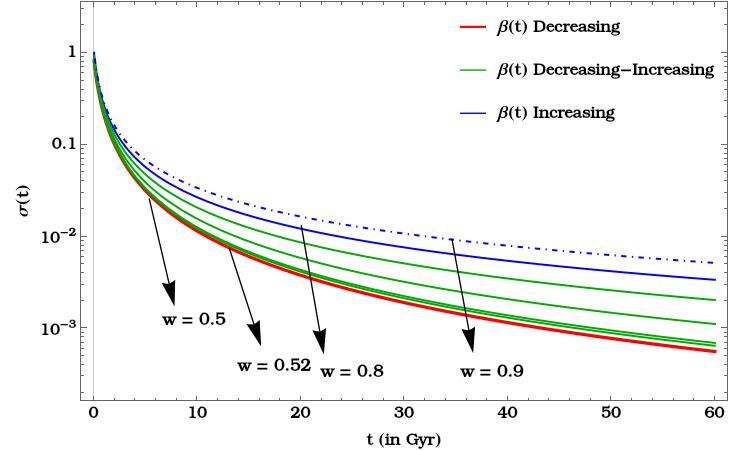} }}
\caption{Evolution of tilt, derivative of tilt, overall scale factor, density, Hubble parameter and shear for constant equations of states with initial conditions $X_{in} = 1$, $Y_{in} = 1$, $\rho_{in} = 0.6$, $\beta_{in} = 1$ at $t=0.01$ Gyr. Note that here we have mostly plotted the cases where $\beta$ shows an increase at late times. To illustrate that initial conditions are important, we have included a case with $w=0.5 >1/3$ where the $\beta$ does not seem to increase, at least during the duration of our evolution. But we also emphasize (as is clear from some of the plots) that the $\beta$ evolution can take a long time to start increasing for some initial conditions.}
    \label{fig:ConstEoSbeta}
\end{figure}

\subsection{Defining Models via \texorpdfstring{${w(t)}$}{w(t)}: Warm up within FLRW}\label{sec:w(t)-FLRW-warmup}

Constant equations of state are not of much interest phenomenologically. This is the reason why in standard cosmology, one considers mixtures of fluids, as we discussed at the beginning of this section.  

From a more phenomenological point of view, what we are doing via allowing mixtures of fluids is to declare that the evolution of the scale factor is dominated by different values of the {\em total} equation of state, at different epochs. Even though not widely stated in this way, this is implicit in the language of the field -- we are all familiar with phrases  like ``radiation dominated epoch'' or ``matter dominated epoch'' or "dark energy dominated epoch". Roughly speaking the {\em total} equation of state behaves as though it is radiation or matter or vacuum energy, depending on the cosmic epoch we are considering. This suggests an alternate (and in the case of FLRW, entirely unnecessary) way to make the Friedmann equations autonomous: we simply pick a function $w_{\text{eff}}(t)$ as our total equation of state, instead of considering mixtures.  This function will be our definition of the ``model''. As we will illustrate, the usual FLRW assumption of a multi-component fluid, each with a constant equation of state, can equivalently be viewed as a specific choice for the {\em total} equation of state $w_{\text{eff}}(t)$. The advantage of the $w_{\text{eff}}(t)$ perspective as opposed to the component fluid approach is that the former generalizes readily to the dipole cosmology setting as well, while the latter does not. 

Let us make the above discussion more concrete in the context of FLRW. We take the cosmic fluid to be composed of perfect fluids with  constant EoS $w_i$. One may describe the cosmological model through the effective EoS $w_{\text{eff}}$,
\begin{equation}\label{w-tot-def}
    w_{\text{eff}}(t):= \frac{\sum_i p_i}{\sum_i \rho_i}=\sum_i w_i\Omega_i, \qquad \Omega_i=\frac{\rho_i}{\sum_i \rho_i},\quad \sum_i \Omega_i=1.
\end{equation}
Recalling the usual Friedmann equations:
\beq\label{Friedmann-Eq}
\sum_i\rho_i=3H^2,\qquad \sum_i p_i=-\frac{\dot{\rho}}{3H}-\rho=-2\dot{H}-3H^2, 
\eeq
we get 
\begin{equation}\label{w-eff-FLRW}
  w_{\text{eff}}(t)=-1-\frac{2\dot{H}}{3H^2}.    \qquad {\text{or}} \qquad  \frac{1}{H(t)}=\frac{1}{H_0}+\frac32\int_{t_0}^t dt\ (w_{\text{eff}}+1)
\end{equation}
$w_{\text{eff}}$ has generically a nontrivial $t$ dependence which determines the geometric quantity $H(t)$.  In our analysis we  can include the cosmological constant term as a component in the cosmic fluid with $w=-1$. For physical systems which respect weak energy condition $0\leq \Omega_i\leq 1$ and for those which satisfy null energy condition $-1\leq w_i\leq 1$. Therefore, $-1\leq w_{\text{eff}}\leq 1$. While $w_{\text{eff}}$ is in general a function of $t$, in different cosmological epochs it may happen that we are in an $i^{\text{th}}$ component dominated era where $w_{\text{eff}}\simeq w_i$ for a specific $i$, e.g. ``radiation dominated'', ``matter dominated'' and ``dark energy dominated'' epochs.

Hereafter, whenever there is no confusion, we drop the subscript ``$\text{eff}$'' and simply write $w(t)$. 
\paragraph{Late-time flat $\Lambda$CDM.} 
Let us make things completely explicit and write down the form of $w(t)$ for flat \lcdm\ after radiation has decoupled. We will use this specific form of $w(t)$ later, in one of our dipole cosmology toy examples (the dipole ``$\Lambda$CDM'' model), where we will use it as a crude model to track quasi-realistic phenomenology in the dipole setting.
  
\begin{figure}[ht]
\centering
\subfloat[\label{fig:La}]{\includegraphics[width = 7.5 cm]{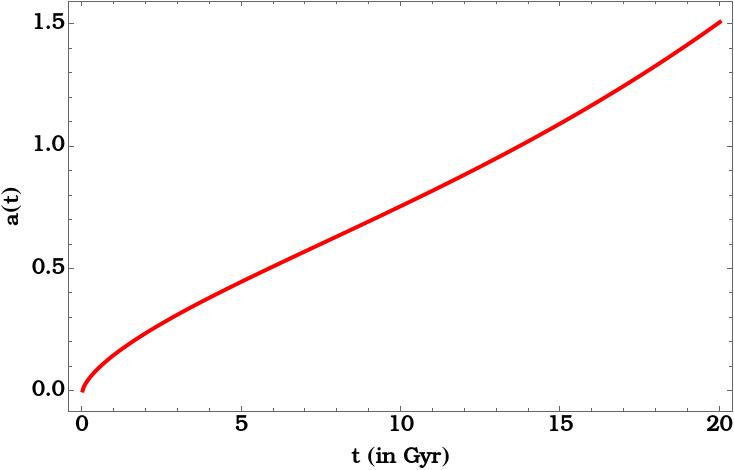}}\hfill
\subfloat[\label{fig:LH}]{\includegraphics[width = 7.5 cm]{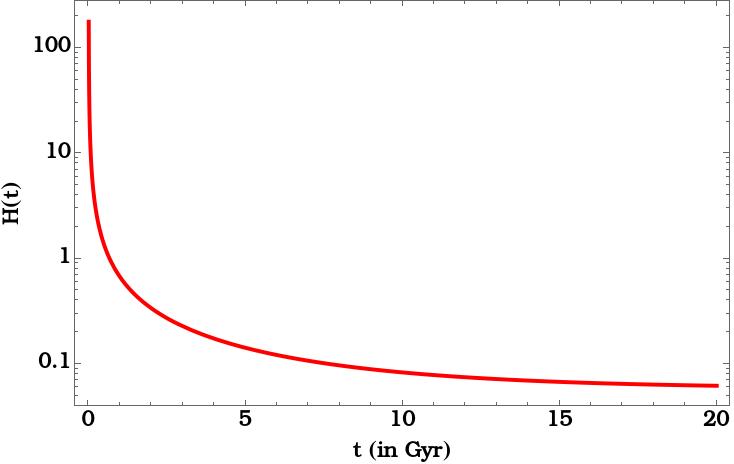}}
\\
\centering
\subfloat[\label{fig:LR}]{\includegraphics[width = 7.5 cm]{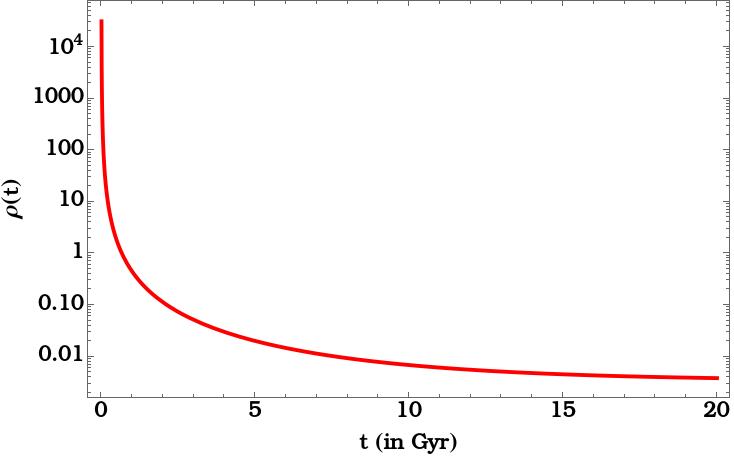}}\hfill
\subfloat[\label{fig:Lw}]{\includegraphics[width = 7.5 cm]{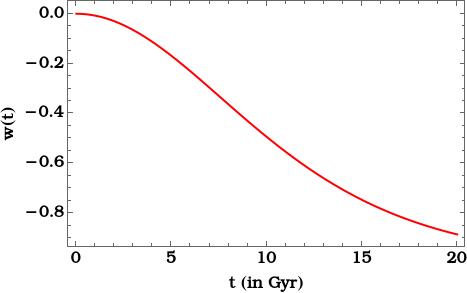}}
\caption{Plots obtained by solving \eqref{eq:LCDM} with typical $\Lambda$CDM parameters. For concreteness, we have used $\Omega_\Lambda=0.6911$ and  $t_0=13.68$ Gyr \cite{Planck-2018}. Figures (\ref{fig:La}), (\ref{fig:LH}), (\ref{fig:LR}), (\ref{fig:Lw}) are plots for the scale factor, Hubble expansion rate, density and equations of state (EoS) respectively. The quantities are plotted from $t = 0.01$ Gyr.  }
\label{fig:avar}
\end{figure}

We can write the Hubble diagram for this set up as 
\bea
H(t)=\frac{\dot{a}}{a}= H_{0}\sqrt{\frac{\Omega_{m0}}{a(t)^{3}} + \Omega_\Lambda}, \qquad  \Omega_\Lambda=1-\Omega_{m0}.\label{eq:LCDM}
\eea
One can then integrate the above equation and find 
\begin{equation}\label{a(t)-LCDM}
a(t)= \left(\frac{\Omega_{m0}}{\Omega_{\Lambda}}\right)^{1/3} 
\sinh^{2/3}\left(\frac32 \sqrt{\Omega_{\Lambda}H_0^2}\ t\right),\qquad H(t)=\sqrt{\Omega_{\Lambda}H_0^2} \coth{\left(\frac32 \sqrt{\Omega_{\Lambda}H_0^2}\ t\right)}.
\end{equation}
The integration constants have been chosen such that  at present epoch $t_0$, $H(t_0)=H_0$, ie. $\sqrt{\Omega_\Lambda} \coth{\left(\frac32 \sqrt{\Omega_{\Lambda}H_0^2}\ t_0\right)}=1$ and $a(t_0)=1$. For  $H_0, \Omega_{m0}$ one may take  the usual Planck values \cite{Planck-2018} where $t_{0}\sim 13.7$Gyr. Using \eqref{w-eff-FLRW} and \eqref{eq:LCDM}, for \lcdm\ one finds
\begin{equation}\label{w-eff-LCDM}
w(t)=-\tanh^2{\left(\frac32 \sqrt{\Omega_{\Lambda}H_0^2}\ t\right)} .
\end{equation}
For far past $H_0t\ll 1$, as expected we are dealing with a matter dominated Universe with $w(t)\sim 0$ and for far future $H_0 t\gg 1$, with a DE (cosmological constant) dominated Universe. In Figure \ref{fig:avar} we present the plots obtained by numerically solving \eqref{eq:LCDM} and \eqref{w-eff-LCDM}, which of course perfectly matches with the analytical expressions given above. 
One can choose to view  $w(t)$ in \eqref{w-eff-LCDM} as the {\em definition} of the flat \lcdm\ model in the post-radiation era. Note that this can also equivalently be viewed as a $\Lambda$-dust model.

\paragraph{$\Lambda$-$w$ models.} We close this section by generalizing the above $\Lambda$-dust expression to a system with cosmological constant $\Lambda$ plus a fluid of constant EoS $w$ ($w\neq -1$):
\begin{subequations}
\begin{align}
    H(t)& = H_{0}\sqrt{\Omega_\Lambda+ \frac{\Omega_{w}}{a(t)^{3(1+w)}} }, \qquad  \Omega_\Lambda=1-\Omega_{w},\label{eq:w-Lambda}\\
a(t) &= \left(\frac{\Omega_{w}}{\Omega_{\Lambda}}\sinh^2(K t)\right)^{\frac{1}{3(1+w)}} ,\qquad H(t)=\sqrt{\Omega_{\Lambda}H_0^2} \coth{(K t)} \label{a(t)-H-Lambda-w}\\
w(t) &=-\tanh^2(K t)+\frac{w}{\cosh^2(Kt)},\qquad K=\frac32 (1+w) H_0\sqrt{\Omega_{\Lambda}}. \label{w-eff-w-Lambda}
\end{align}
\end{subequations}
Note that for far future, $Kt\gg 1$, $w(t)\simeq -1+4(1+w) e^{-2Kt}$ and $H(t)\simeq \sqrt{\Omega_{\Lambda}H_0^2}$. The fact that the rate at which the total EoS falls, increases with the stiffness $w$ will play a role in some of our discussions in the more general context of dipole cosmology as well.

\subsection{Defining Models via \texorpdfstring{${w(t)}$}{w(t)}: Dipole Cosmology}

The above discussion of models in the FLRW setting serves as a natural jumping off point for a similar discussion in Dipole Cosmology. Defining models via choosing $w(t)$ is the setting in which we will explore the possibility of quasi-realistic cosmologies in the dipole context. 

But what about dipole model building where there are multiple fluid components? After all, this is the more familiar setting for model-building in FLRW. It turns out that this is {\em not} meaningful in Dipole Cosmology, because all the fluid components share the same tilt. In other words, if we had a fluid component with the equation of state $w_i$ that was individually covariantly conserved, we could re-arrange the analogues of eqns. \eqref{Con1} and \eqref{Con2} to get 
\bea
w_i \times \Big(\frac{\dot{X}}{X}+2\frac{\dot{Y}}{Y}+\frac{d}{dt} \log \cosh \beta-\frac{2}{X}\tanh \beta\Big)=\Big(\frac{\dot{X}}{X}+\frac{d}{dt} \log \sinh \beta\Big)
\eea
But this equation has to apply for {\em each} of the components, which is only possible if each of the equations of state $w_i$ were the same. One might consider relaxing the possibility that each component has to be separately conserved and simply demand that the total stress tensor is conserved. But then we are left with  $X(t), Y(t), p_i(t) ,\rho_i(t)$ and $\beta(t)$ as the independent variables with $i$ ranging over (say) $N$ components. The equations of state  $w_i$ will reduce the total number of degrees of freedom from $3+2 N$ to $3+N$. But with only the four ODEs of the original Dipole Friedmann system, the system is not autonomous, if we allow multiple components (ie., $N >1)$). 

In other words, the advantage of the $w(t)$ perspective, as opposed to the component fluid approach, is that the former can be generalized  to the Dipole Cosmology setting, while the latter does not. Because of this, in order to get a sense of the dynamics of Universes with flows, we will use the $w(t)$-based approach. Let us emphasize however two caveats. Choosing $w(t)$, while pragmatically useful, is {\em far} from a minimal approach to model building. This is because we are introducing a {\em function} as opposed to a few parameters, so without a further rationale for restricting the form of the function, this is a rather extravagant approach. In later sections, we will choose simple functions with only a few parameters in them for illustration -- there may be better choices for these functions that are more physically motivated. We will be interested in robust features that are not too dependent on the specifics of the function, and our goal is to extract some general messages regarding Universes with flows. So these issues will not be a problem for us in the present paper. But detailed model building in Universes with flows, will have to contend with these challenges.

The general argument against multi-component fluids has an important and interesting exception: for the $w=-1$ case, the tilt parameter drops out of the dipole cosmology equations. That is, $\beta$ can take any arbitrary value for this case. Therefore, one is allowed to consider a generic cosmic fluid together with a cosmological constant. Indeed, the dipole equations of the previous section were written for such a general case, where we had a generic tilted fluid plus a cosmological constant. This in particular means we can have dipole $\Lambda$-dust or dipole $\Lambda$-radiation models (for example). 

In the following, we will consider various models defined via suitably chosen functions for $w(t)$. These examples are shown for illustrative purposes -- we are mainly interested in generic and robust features that are not too dependent on the specifics of the chosen model. Our goal is to extract some general lessons regarding Universes with flows, rather than detailed model building. 
We will be particularly interested in cases that are broadly similar to our Universe, e.g. $w(t)$ asymptotically approaching $w= -1$ in future. For these examples which yield an accelerated expansion at late times, we find that the dipole flow $\beta$ can be increasing in the accelerating phase even though the shear anisotropy $\sigma$  is decreasing.

\subsection{Flows Can Grow Even in Accelerating Models}

One of our goals in this paper is to see if there are models which have loosely similar evolution histories for the scale factor as flat LCDM. The caveat is that we would like to have this, while at the same time allowing dipole flows (captured by $\beta$) that do {\em not} die down with time. As just noted, we find that this can happen when the {\em total} equation of state  $\rightarrow -1$ at late times\footnote{The models we consider in this section have $w(t) \rightarrow -1$, but do not contain an explicit cosmological constant. In the next section, we will discuss examples where there is an explicit cosmological constant coupled to matter with $w > -1$, which also allow increasing $\beta$ at late times.}. Interestingly, this also ensures late time acceleration! Even though shear dies down in an accelerating Universe as is expected from various pieces of folklore, the fluid flow $\beta$ can increase, is one of the messages we would like to emphasize in this paper. 

We will illustrate this explicitly for three examples where $w(t)$ decrease with time and tend to $-1$ at late times. In all three cases, we find that $\beta(t)$ can increase even when the shear is dying down. The first example we choose is the $w(t)$ that we found earlier for the flat LCDM model, namely \eqref{w-eff-LCDM}. Note that the total $w(t)$ is decreasing in flat LCDM. When we use this $w(t)$ in the context of Dipole Cosmology, we will call this the dipole ``$\Lambda$CDM'' model. The quotes are in place to emphasize that the equation of state is simply borrowed {\em ad-hoc} from \eqref{w-eff-LCDM} as a quasi-realistic phenomenological model that exhibits late-time acceleration. In particular, the model should not be confused with the dipole $\Lambda$-dust model.

\begin{figure}[H]
    \centering
    \subfloat[\label{fig:wL}]{{\includegraphics[width=8.0 cm]{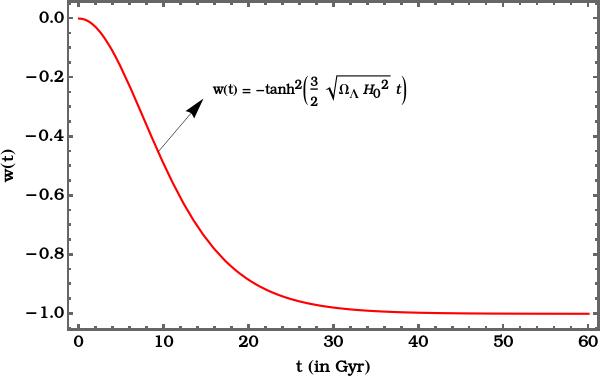} }}\hfill
    \subfloat[\label{fig:aVarL}]{{\includegraphics[width=8.0 cm]{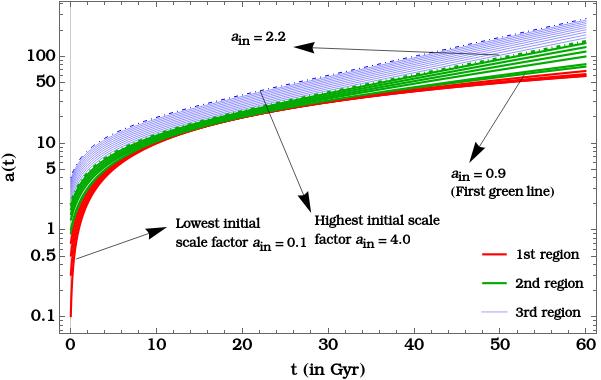} }}\\
     \subfloat[\label{fig:DLCDMH}]{{\includegraphics[width=8.0 cm]{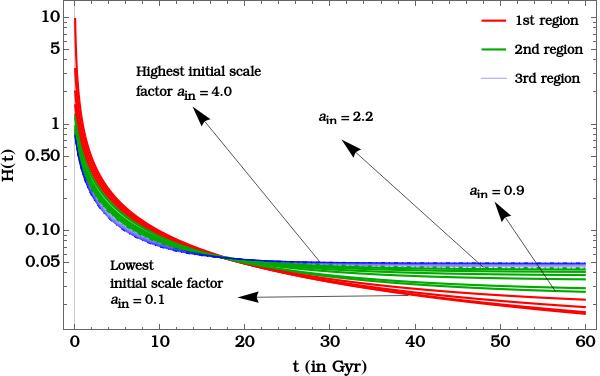} }}\hfill \subfloat[\label{fig:rhoVarL}]{{\includegraphics[width=8.0 cm]{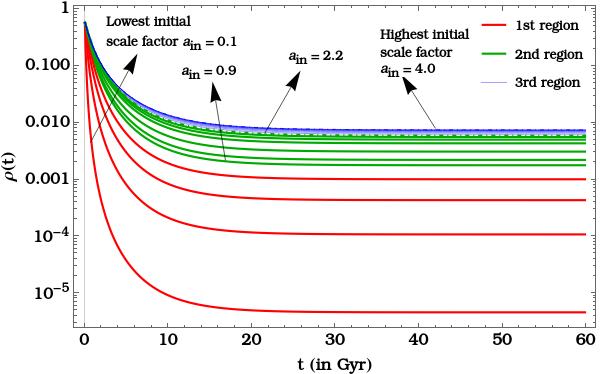} }}\\ 
      \subfloat[\label{fig:betVarL}]{{\includegraphics[width=8.0 cm]{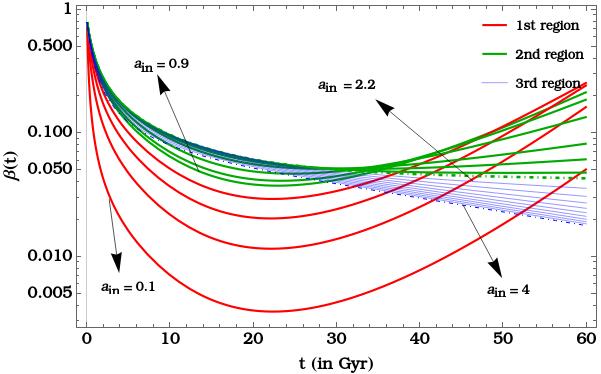} }}\hfill
      \subfloat[\label{fig:sigmaVarL}]{{\includegraphics[width=8.0 cm]{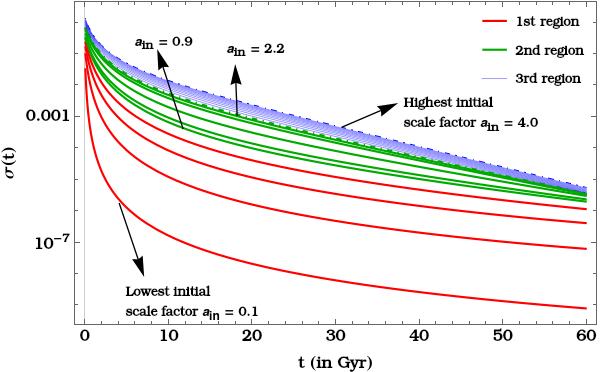} }}
\caption{Time evolution of the dipole ``\lcdm'' model. This model is defined by $w(t)$ given in plot (\ref{fig:wL}) which is the same as the one in plot (\ref{fig:Lw}). 
The initial conditions are $\rho_{in}=0.6, \beta_{in}=0.793, b_{in}=0$ at $t=0.01$ Gyr. While the shear $\sigma$ dies off fast, the tilt $\beta$ remains relatively large, and depending on the initial conditions can also grow in time. For some of the initial conditions, the time axes in our plots are not long enough to see that the Hubble parameter stabilizes to a constant value and that the drop in $\sigma$ is exponential. But we have checked that these indeed happen. We have also checked that at late times, $\sigma \sim a^{-3}$. }
    \label{fig:Dipole-LCDM}
\end{figure}

The plots in Figure \ref{fig:Dipole-LCDM} may be compared to those of usual flat \lcdm\ (cf. Figure \ref{fig:avar}) or those of dipole-dust (cf. Figure \ref{fig:ConstantEOS}, $w=0$ plots). While the anisotropy (shear) drops quite fast and we tend to a usual FLRW metric, the tilt need not die down and can even increase. Also compared to the dipole-dust case with $\Lambda=0$, $\sigma$ drops (much) faster in the dipole ``$\Lambda$CDM'' case. This is unsurprising: As Wald's cosmic no-hair theorem indicates, anisotropy should die off exponentially fast in the presence of a positive cosmological constant. The non-trivial and remarkable result here is that the bulk flow $\beta$ does not trace the shear $\sigma$. In the plots, we have focused on initial conditions that are fairly isotropic (we often work with $X_{in}=Y_{in}$) because of our goal is to see the evolution of shear and tilt near such configurations. Note that $X_{in}=Y_{in}$ does not mean that the initial shear has to be zero, because the latter also contains derivative information. 

\newpage

\begin{figure}[H]
    \centering
    \subfloat[\label{fig:w2}]{{\includegraphics[width=8.0 cm]{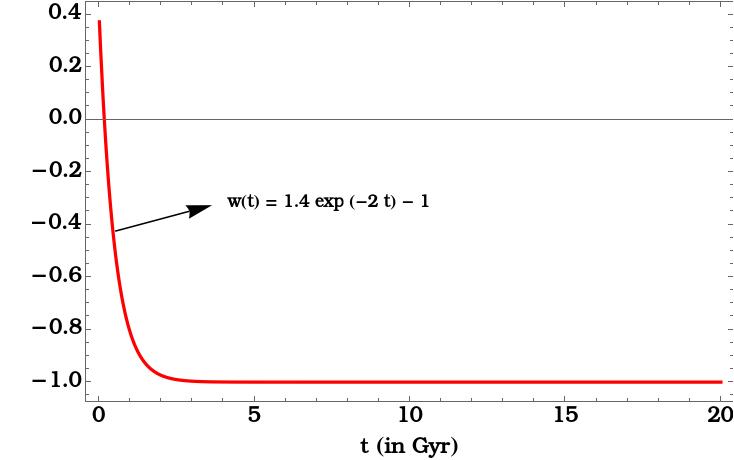} }}\hfill 
    \subfloat[\label{fig:aVar2}]{{\includegraphics[width=8.0 cm]{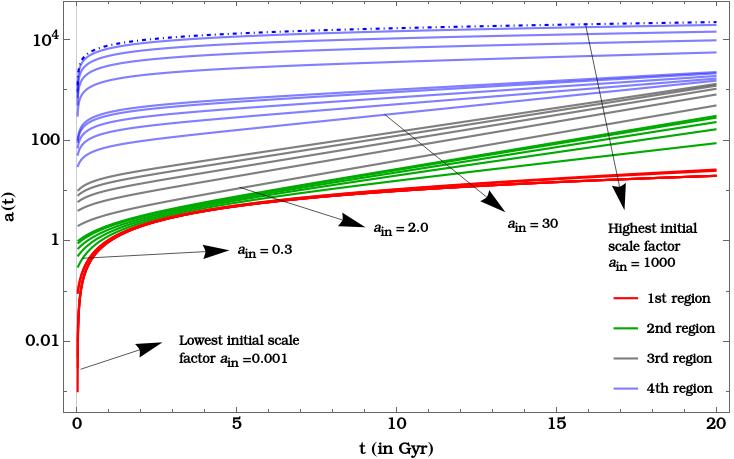} }}\\ \subfloat[\label{fig:w2h}]{{\includegraphics[width=8.0 cm]{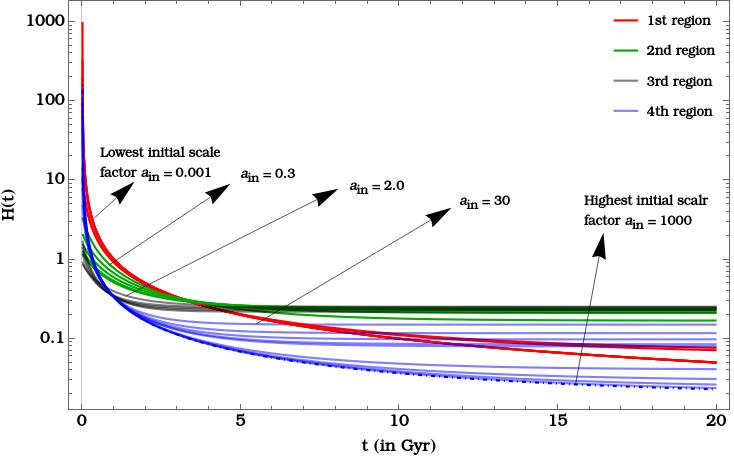} }}\hfill
     \subfloat[\label{fig:rhoVar2}]{{\includegraphics[width=8.0 cm]{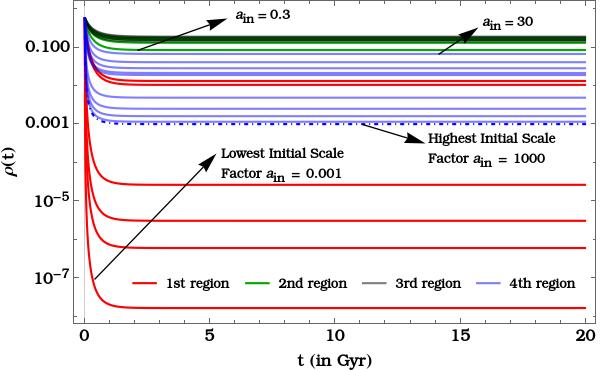} }}\\ 
      \subfloat[\label{fig:betVar2}]{{\includegraphics[width=8.0 cm]{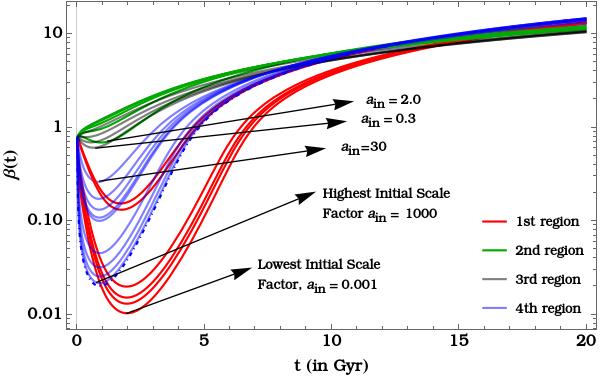} }}\hfill
      \subfloat[\label{fig:sigmaVar2}]{{\includegraphics[width=8.0 cm]{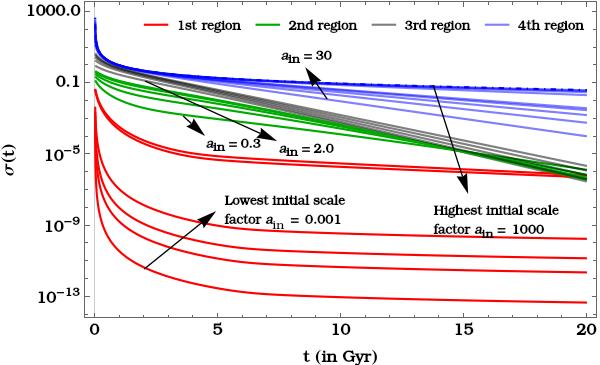} }}
\caption{Time evolution of the model defined with $w(t)$ given in plot (\ref{fig:w2}). The initial conditions are the same as in the previous figure. While the anisotropic shear $\sigma$ is a monotonically decreasing function, the tilt $\beta$ can be increasing at late times. As in the previous figure, for some of the initial conditions, the time axes in our plots are not long enough to see that the Hubble parameter stabilizes to a constant value and that the drop in $\sigma$ is exponential. But we have checked that these indeed happen. We have also checked that at late times, $\sigma \sim a^{-3}$. }
    \label{fig:Dipole-w-exponential}
\end{figure}
\begin{figure}[H]
\centering
\subfloat[\label{fig:Varw}]{\includegraphics[width = 7.5 cm]{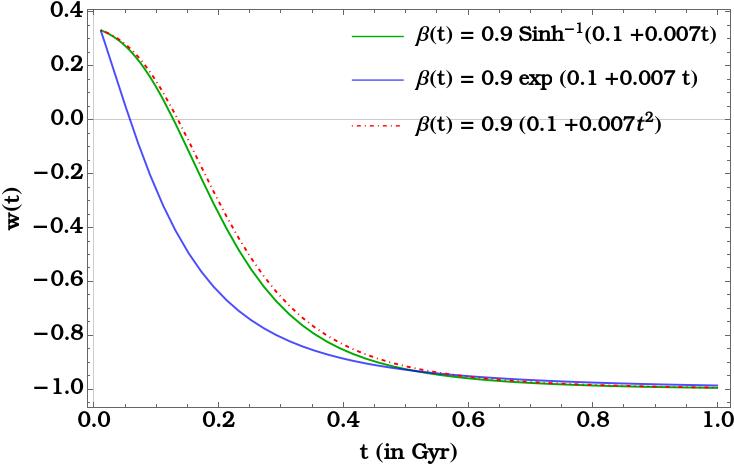}}\hfill 
\subfloat[\label{fig:Vara}]{\includegraphics[width = 7.5 cm]{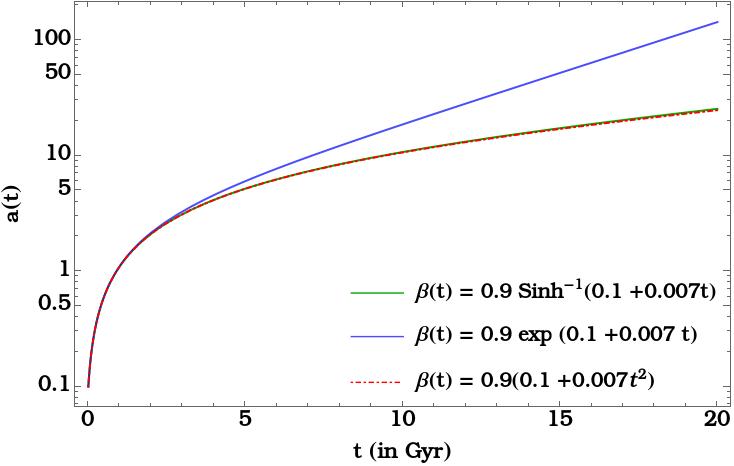}}\\
\subfloat[\label{fig:Varb}]{\includegraphics[width = 7.5 cm]{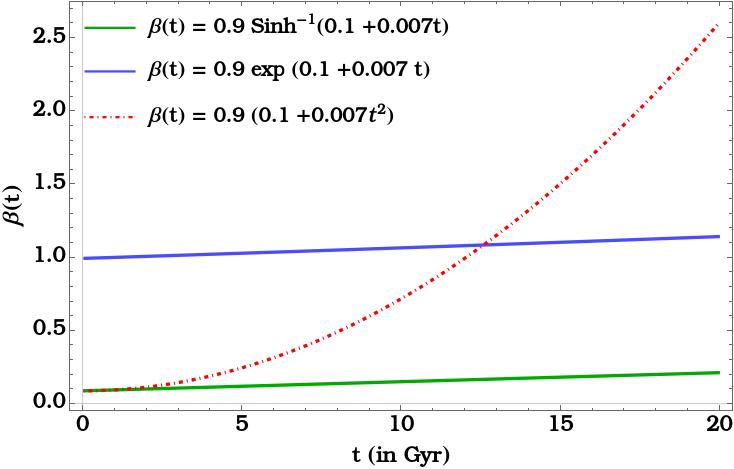}}\hfill
\subfloat[\label{fig:Vars}]{\includegraphics[width = 7.5 cm]{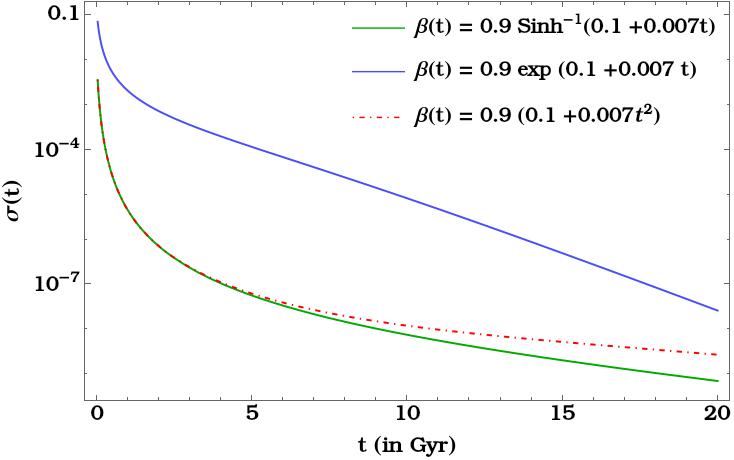}}
\caption{Plots of $w(t), a(t)$, $\beta(t)$, $\sigma(t)$ for the system of equations \eqref{FW1}, \eqref{FW2}, \eqref{Con1}, \eqref{Con2} for $\Lambda=0$ and $\beta(t)$'s defined in Figure (\ref{fig:Varb}). The initial conditions are fixed at $\rho_{in} = 0.6$, $p_{in} = 0.2$, $a_{in} = 0.1, b_{in}=0$ at $t=0.01$ Gyr. Note that we need an initial condition in pressure because we are fixing the functional form of $\beta(t)$ instead of $w(t)$. The scale factor exhibits acceleration. }
\label{fig:Varaious-beta(t)}
\end{figure}

Apart from dipole ``$\Lambda$CDM'' model, we consider two classes of examples. We first consider  an exponentially decreasing $w(t)$ that tends to $-1$. The results of the evolution of the system is depicted in Figure \ref{fig:Dipole-w-exponential}. Our third class of examples is obtained by demanding that $\beta(t)$ has some (specific) increasing form. In other words, for this class of examples, to render dipole cosmology equations autonomous, instead of $w(t)$, we specify $\beta(t)$  to have a given form. We then evolve the system with this chosen $\beta(t)$. This is somewhat like reverse-engineering the model. The results of this analysis is shown in Figure \ref{fig:Varaious-beta(t)}. In various cases we have tried (which include linear, power law and exponential growths of $\beta$), the equation of state decreases with time and the Universe accelerates. 


These observations are robust, but numerical. It will be good to have an analytical demonstration/understanding of these facts. This is one of the things we will do in the next section. We will set up a perturbation theory for our equations of motion around a late time limit, under the assumption that $w(t)$ goes to $-1$ at late times. In this limit, we can do a perturbative expansion for the Dipole Cosmology equations and the system simplifies at leading orders. We will be able to make various (semi-)analytic observations and note that $\beta$ growth at late times is quite generic in late time accelerating dipole cosmologies. 

To conclude this section -- We saw tilt growth in multiple accelerating dipole cosmologies. But let us emphasize that one can have tilt growth in decelerating cosmologies as well, as we noted earlier in examples with constant equation of state with $w > 1/3$. We also re-iterate that the results we found for the accelerating Universes here are not at odds with Wald's cosmic no-hair theorem -- the theorem only focuses on the metric and that shear goes to zero, whereas here we focus on tilt $\beta$ which is a property of the fluid.



\section{A More Detailed Analysis}\label{sec:5}

In the previous section, we found multiple (numerical) examples of increasing $\beta$ for various choices of EoS that asymptote to $-1$ at late times. In  this section, we will put the numerical solutions of the last section on a more firm footing by doing an analytic perturbative analysis of the $w(t)$ goes to $-1$ set up. First, in subsection \ref{sec:pert-EoM} we set up the perturbation theory of the field equations to first order. Then in section \ref{sec:tilt-growth}, we will identify a simple set of conditions under which tilt can grow at late times in an accelerating Universe. In section \ref{sec:growth-generacity}, we consider the specific scenario where the $w(t\rightarrow\infty)=-1$ condition on the total effective EoS is realized via a positive cosmological constant. In this case, under some further minor assumptions, we show that the condition identified in $\ref{sec:tilt-growth}$ translates to a statement about the EoS of the rest of the matter, if $\beta$ is to increase at late times. In section $\ref{sec:flowingdark}$, we emphasize that the $w(t\rightarrow\infty)=-1$ condition is more generic than the cosmological constant scenario discussed in \ref{sec:growth-generacity}. We call this more general scenario, $flowing$ $dark$ $energy$. Finally in section \ref{sec:away-from-far-future} we briefly discuss situations where the tilt $\beta$ increases at intermediate times before dying down eventually -- this of course is independent of the perturbative analysis of \ref{sec:pert-EoM}. The primary message of this section is that increasing $\beta$ is easy enough to arrange in an expanding (even accelerating) Universe.



\subsection{Perturbation Theory}\label{sec:pert-EoM}

To set up the perturbative analysis we take our EoS to be 
\begin{equation}
    w(t) = -1 + \epsilon\,\Omega(t),
\end{equation}
where we will view $\Omega(t)$ as a positive, decreasing function that limits to zero at late times.  
{One may perform the above perturbative analysis in terms of $a(t), b(t)$ functions 
\begin{equation}
a(t)=a_0(t) (1+\epsilon a_1(t)),\qquad \rho=\rho_0(1+\epsilon \rho_1(t)), \qquad b(t)=b_0(t)+\epsilon b_1(t),
\end{equation}
which yields, 
\begin{equation}
H(t)=H_0(t) +\epsilon \dot{a}_1(t),\qquad \sigma=\sigma_0+\epsilon\sigma_1=3\dot{b}_0(t)+3\epsilon \dot{b}_1(t).
\end{equation}
Instead of $a$ and $b$ variables, we can also do the perturbative analysis in terms of $X$ and $Y$ variables. The latter is simpler technically (we present some details in Appendix \ref{sec:appendix-pert}), but the former has the advantage of being more directly connected to quantities like shear and expansion. We have analytically checked that the two sets of perturbative equations are mutually consistent. Our exact numerical evolution, which were done in the $X, Y$ language, have been checked directly against the perturbative results in the $X, Y$ variables at late times.

Let us first analyse the lowest $\mathcal{O}(\epsilon^0)$ equations. Eq.(\ref{EoM-H-sigma-d}) yields $\sigma_0=0$ or $b_0=$constant. This constant $b_0$ may be set to 0 upon a rescaling of $x,y$ coordinates and $a_0$. Eq.\eqref{Con1-1} yields $\rho_0=$constant and \eqref{EoM-H-sigma-a}, 
\begin{equation}\label{a0-H0-perturbative}
 a_0(t)= \frac{a_0}{{\cal H}_{\infty}} \sinh({\cal H}_{\infty}t),\qquad  H_0(t)= {\cal H}_{\infty}\coth({\cal H}_{\infty}t), 
\end{equation}
where 
\begin{equation}
    {\cal H}_{\infty} :=\sqrt{\frac{\rho_0}{3}}
\end{equation}
is the asymptotic Hubble expansion rate, ie., $H_0(t)$ goes to the constant ${\cal H}_{\infty}$ at late times (${\cal H}_{\infty} t\gg 1$). The $\mathcal{O}(\epsilon^0)$ equations above have de Sitter evolution, but it is crucial to note that there need not be an explicit cosmological constant here. In fact, as we have already seen in the plots in the previous section, the asymptotic value of $\rho$ at late times can vary depending on the initial conditions, and is not a property of the dipole cosmology equations. This means that ${\cal H}_{\infty}$ is an asymptotic property of a given solution, and not a property of the theory. As we will see later, this is because $w(t)\rightarrow -1$ need not arise from a cosmological constant, and can arise from more  general (flowing) dark energy.

Let us also note that $\beta_0$ is not fixed at $\mathcal{O}(\epsilon^0)$, as expected: since $\rho+p= 0$ at lowest order, $\beta_0$ remains unspecified at this order. The zeroth order solutions are hence specified by a single constant $\rho_0$ or ${\cal H}_{\infty}$. 

Next, we explore order $\mathcal{O}(\epsilon^1)$ equations. Eq.\eqref{EoM-H-sigma-d} yields
\begin{equation}\label{sigma1}
    \sigma_1= \frac{3{\cal H}_{\infty}}{4}\, \Omega \, \sinh(2\beta_0) \sinh ({\cal H}_{\infty}t)
\end{equation}
and \eqref{EoM-H-sigma-a}, \eqref{Con1-1}, \eqref{Con2-1} yield
\begin{subequations}
\begin{align}
&\frac{d}{dt}\left(\tanh({\cal H}_{\infty}t) \ a_1\right)+2{\cal H}_{\infty}b_1=\frac{{\cal H}_{\infty}}{2}\tanh^2({\cal H}_{\infty}t)(\rho_1+\Omega \sinh^2\beta_0+4b_1) \label{a1-dot}\\
&\dot{\beta}_0= \frac12\tanh{2\beta_0}\left(\Delta(t)+ 2{\cal H}_{\infty}\frac{\tanh\beta_0}{\sinh({\cal H}_{\infty}t)}\right)  \label{beta0} \\
&\dot{\rho}_1=-\frac12\Omega\tanh{2\beta_0}\left[{\Delta(t)}\tanh\beta_0 + \frac{2{\cal H}_{\infty}}{\sinh({\cal H}_{\infty}t)} \left(3\coth{2\beta_0} \cosh({\cal H}_{\infty}t)-  1\right)\right],\label{rho1}
 \end{align}
\end{subequations}
where 
\begin{equation}\label{Delta(t)}
   \Delta(t):= -\frac{\dot{\Omega}}{\Omega}-4H_0(t)=-\frac{1}{a_0(t)^4}\frac{d}{dt}\ln(\Omega a_0(t)^4).
\end{equation}
\eqref{shear-EoM} does not yield an independent equation. Therefore, we have 4 equations for $\beta_0, a_1,\rho_1,\sigma_1$ which may be solved in terms of $\Omega(t)$.}

\subsection{Can Flows 
Grow at Late Times?}\label{sec:tilt-growth}

One of the main questions we would like to explore is the late time behavior of $\beta$ (or $\beta_0$ in perturbation theory). In particular, we want to know if at late time (${\cal H}_{\infty} t\gg 1$)  $\beta_0$  can be increasing. To this end, let us analyse \eqref{beta0} more closely. The last term on the RHS, is positive, but falls off exponentially fast in time and one may ignore it at late times. Let us note also that because of the symmetries of the dipole cosmology system, we can always work with positive $\beta$ without loss of generality. The sign of $\dot{\beta}_0$ is essentially determined by the sign of $\Delta(t)$; if $\Delta(t)$ is positive (negative), we get a growing (decreasing) $\beta_0$. For 
$\Delta=0$ the higher order terms and the other exponentially dying off term which we have neglected will determine if $\beta_0$ is growing or decreasing. 
\begin{figure}[H]
    \centering
    \includegraphics[height = 7.5 cm, width = 12 cm]{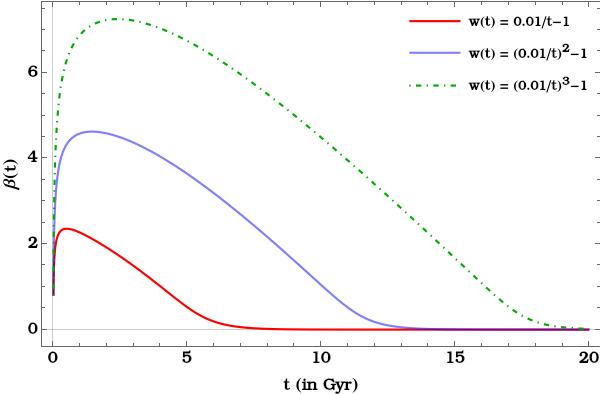}
    \caption{Sample tilt plots for $\Omega \propto t^{-\gamma}$.  We see from the plots that as $|\gamma|$ increases the maximum of $\beta(t)$ shifts to later times. The initial conditions are $a_{in}=0.8, \beta_{in}=0.793, \rho_{in}=0.6$ at initial time $t=0.01$ Gyr.}
    \label{fig:gamTilt}
\end{figure}

\begin{figure}[H]
    \centering
    \subfloat[\label{fig:betaPert}]{{\includegraphics[width=8.0 cm]{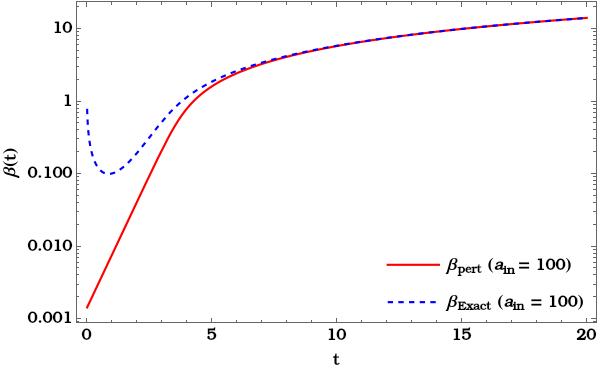} }}\hfill
    \subfloat[\label{fig:sigmaPert}]{{\includegraphics[width=8.0 cm]{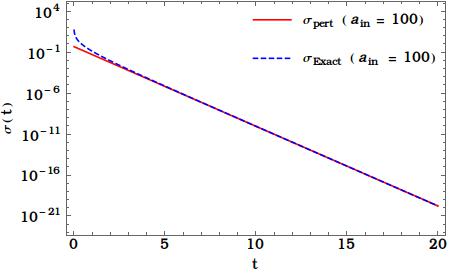} }}

    \caption{Example of matching between exact numerical solution and the late time perturbative solution for $\beta$ and $\sigma$, with the choice $\Omega(t) = 1.4 \exp{(-2t)}$. The initial conditions are $X_{in} = 100, Y_{in}=100, \rho_{in}=0.6, \beta_{in}=0.793$ at $t=0.01$ Gyr. We use the late time values of the exact curves to fix the constants in the perturbative solution, and this matches the curves. }
    \label{fig-pert-match}
\end{figure}

At late times, $H_0(t)$ goes to ${\cal H}_{\infty}$. Therefore, if $\frac{\dot{\Omega}}{\Omega}+4{\cal H}_{\infty}\leq 0$, $\beta_0$ can grow. If $\Omega\sim t^\gamma$ at late times, $\beta_0$ can grow if $\gamma+4{\cal H}_{\infty}t\leq 0$. This condition cannot be met with a finite $\gamma$ at late times. Another generic choice is $\Omega\sim e^{-\alpha t}$. Apart from being of general interest, we will see in the next section that the exponential form arises very naturally in a Universe with a positive cosmological constant. 
In these cases, the tilt grows if $\alpha\geq 4{\cal H}_{\infty}$. Explicitly, one can integrate \eqref{beta0} at late times to get
\begin{equation}\label{late-time-beta0}
    \sinh 2\beta_0 \simeq C e^{(\alpha-4{\cal H}_{\infty})t},
\end{equation}
where $C$ is an integration constant. So $\beta_0(t)$ can grow linearly in time, though with a different pace than $H_0(t)$. On the other hand, \eqref{rho1} shows that at late times $\dot{\rho}_1 <0$ and that $\rho_1$ decreases with an exponential fall off (as $\sim e^{-\alpha t}$) at late times. Similarly, \eqref{a1-dot} indicates that $a_1$ has an exponential fall off (with the smaller of $\alpha$ or $4{\cal H}_{\infty}$). Finally, \eqref{sigma1} implies that $\sigma_1\sim a_0^{-3}$, irrespective of $\alpha$. This is compatible with the usual lore that shear in the presence of an exponential growth dies off with power $-3$ of the scale factor. Note that in making some of the claims above, we have set the integration constants in some of the first order quantities to zero. This follows from our expectation that perturbation theory is an increasingly good approximation at late times; the zeroth order solution must be valid at very late times\footnote{We can also view this as absorbing the first order integration constants into the zeroth order ones, while matching with the exact solutions.}. We have verified that the full evolution match extremely well with perturbation theory, for specific choices of $\Omega$. See Figure \ref{fig-pert-match}. Note also that the fact that $\beta$ decouples at late times is crucial for some of these discussions.

Let us summarize some of the salient features we observed. Our perturbative analysis indicates that if $\Omega$ has a power-law fall-off at late times, $\beta_0$ cannot grow. On the other hand, if it has an exponentially decreasing form $\Omega\sim e^{-\alpha t}$,  it can grow depending on the sign of $\alpha-4{\cal H}_{\infty}$. The geometry and the energy density asymptote to their final values exponentially fast. 



\begin{figure}[H]
    \centering
    \subfloat[\label{fig:alphaH-dipoleLCDM-1}]{{\includegraphics[width=8.0 cm]{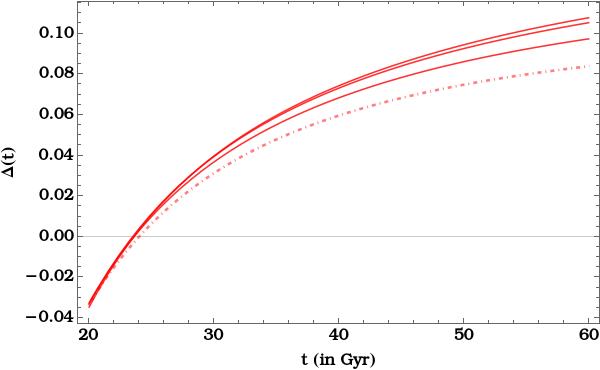} }}\hfill
    \subfloat[\label{fig:beta-dipoleLCDM-1}]{{\includegraphics[width=8.0 cm]{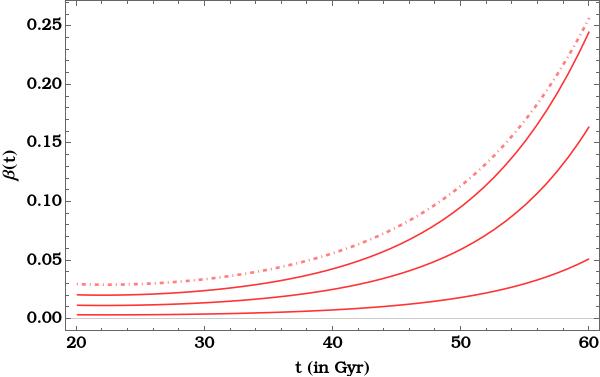} }}\\
    \subfloat[\label{fig:alphaH-dipoleLCDM-2}]{{\includegraphics[width=8.0 cm]{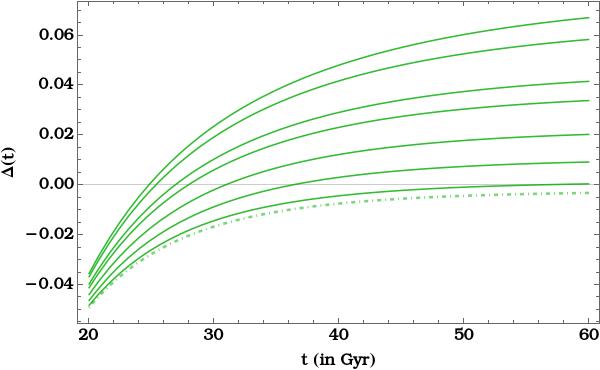} }}\hfill
    \subfloat[\label{fig:beta-dipoleLCDM-2}]{{\includegraphics[width=8.0 cm]{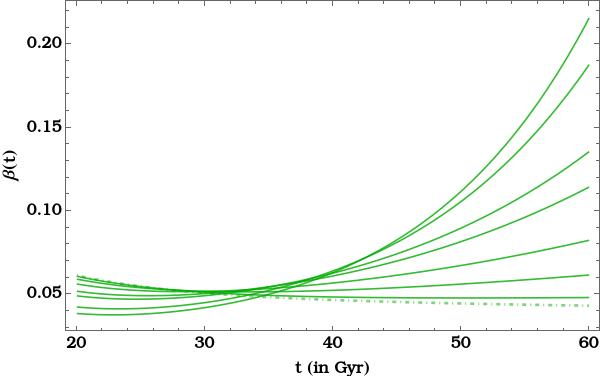} }}\\
    \subfloat[\label{fig:alphaH-dipoleLCDM-3}]{{\includegraphics[width=8.0 cm]{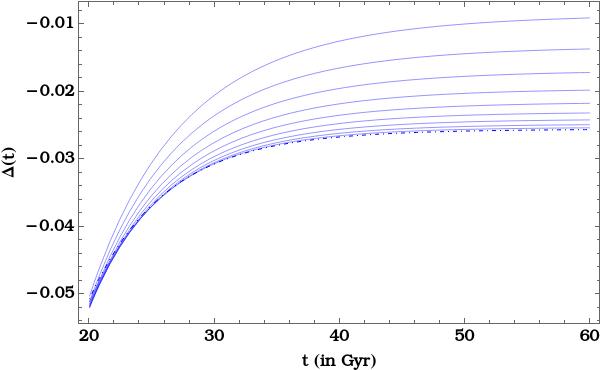} }}\hfill
    \subfloat[\label{fig:beta-dipoleLCDM-3}]{{\includegraphics[width=8.0 cm]{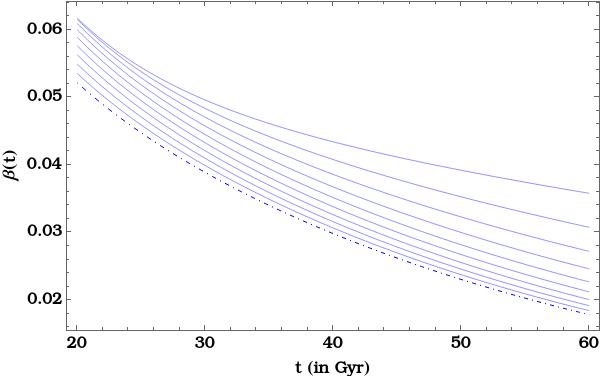} }}
\caption{The plots of $\Delta(t)$ and $\beta$ for dipole ``\lcdm'' model for different regions in the late time parameter space., under exact numerical evolution. This illustrates the validity of the perturbative results for this example. The color-coding is consistent with the color-coding in Figure \ref{fig:Dipole-LCDM}.}
    \label{fig:dipole-LCDM-beta-growth}
\end{figure}

For  dipole ``\lcdm" model,  \eqref{w-eff-LCDM} indicates $\Omega\sim 4e^{-3\sqrt{\Omega_\Lambda H_0^2} t}$ i.e. $\alpha= 3\sqrt{\Omega_\Lambda H_0^2}$. The late time behavior of $H_0(t)$ in \eqref{a0-H0-perturbative} and  $H(t)$ in \eqref{a(t)-LCDM} are not necessarily the same, ie. ${\cal H}_{\infty}$ and $\alpha$ are independent parameters. This is because even though we are working with \eqref{w-eff-LCDM} as the EoS, the system we are considering is dipole cosmology, and not the FLRW system. So ${\cal H}_{\infty}$ is controlled by the initial conditions of the full dipole cosmology equations. Depending on the value of ${\cal H}_{\infty}$ the $\beta_0$ growth condition may or may {not} be  satisfied; a fact we are already familiar with from the plots in the previous section. In Figure \ref{fig:dipole-LCDM-beta-growth} we have plotted $\beta$ as a function of $t$ for some different values of the late time parameter. These are exact evolution plots, but they clearly illustrate the transition from decreasing to increasing $\beta$ as the sign of $\Delta$ changes. 

\subsection{Growing Flows and the Cosmological Constant}
\label{sec:growth-generacity}

In the previous subsection we discussed the late time behavior of dipole cosmology with a $total$ EoS $w(t)$ which tends to $-1$ at late times. We found that when $1+w(t)$ has a fast enough exponential fall off,  tilt can grow in time. In this subsection we will show that this sort of scenario arises automatically in models with a cosmological constant, and determine a condition on the late-time EoS of the rest of the matter (ie., except $\Lambda$) so that $\beta$ increases at late times. In other words, we consider examples consisting of a positive cosmological constant $\Lambda$ together with another fluid of density $\rho(t)$ and EoS $w(t)$.  From the FLRW discussion in section \ref{sec:w(t)-FLRW-warmup}, we saw that  the constant-$w$ cosmologies of this type show exponential fall off in the $total$ effective EoS towards $-1$. Indeed,  the larger the $w$, the faster the fall off. It is plausible that a similar feature may happen in dipole cosmology as well, and therefore from our perturbative analysis they are good candidates for growing-$\beta$ cosmologies. We will see that this expectation is indeed correct.

Consider  a dipole cosmology consisting of a fluid with EoS $w(t)$ and energy density $\rho(t)$  and add a cosmological constant $\Lambda$ to it\footnote{In this and the next subsection, we will use the notation $w_{eff}(t)$ to denote the total effective EoS (that includes the cosmological constant). Elsewhere, we will simply use $w(t)$ to denote the total EoS (typically in situations where there is no explicit cosmological constant).}. Here $-1<w(t)\le 1$. This system is described by an effective EoS $w_{eff}(t)$, as can be seen by absorbing the $\Lambda$ into the total tilted fluid stress tensor:
\begin{equation}
w_{eff}(t)=-1+\frac{1+w(t)}{1+\frac{\Lambda}{\rho(t)}}. \label{shiftL}
\end{equation} 
We will assume that $w(t)$ goes to a constant $w_\infty$ at late times, and also that the overall scale factor $a(t)$ has an exponential growth $a(t)\sim e^{{\cal H}_\infty t}$, where ${\cal H}_\infty=\sqrt{\Lambda/3}$. This can be loosely viewed as a consequence of the cosmic no-hair theorem. We will not prove it for our setting, but we will see that this is a self-consistent assumption in what follows. We have also checked this by explicit numerical evolution for $w(t)=w$ (for various constant values of $w$), even though the details will be omitted here. 


The conservation equations for the dipole $\Lambda$-$w(t)$ system at late times (with the assumption that the EoS of the fluid saturates to a constant $w_{\infty}$) take the form	
\bea
		\dot{\rho} + 3 H_{\infty}\rho(1 + w_{\infty}) = -\rho(1 + w_{\infty})\tanh{\beta}\dot{\beta} \label{laterho} \\
		w_{\infty}\dot{\rho} + H_{\infty}\rho(1 + w_{\infty}) = -\rho(1 + w_{\infty})\coth{\beta}\dot{\beta} \label{latep}
\eea
where we have dropped terms that die down faster at late times. Decoupling into $\dot{\rho}$ and $\dot{\beta}$ equations we get
\begin{subequations}
	\begin{align}
	\frac{\dot{\rho}}{\rho} &=-3H_{\infty}(1+w_{\infty})\Bigg(\frac{\coth{\beta}-\frac{1}{3}\tanh{\beta}}{\coth{\beta}-w_{\infty}\tanh{\beta}}\Bigg)\label{DerDensity}\\
	\dot{\beta} &= \frac{(3w_{\infty}-1)H_{\infty}}{\coth{\beta}-w_{\infty}\tanh{\beta}} \label{DerBeta}
	\end{align}	
	\label{DerEq}
\end{subequations} 
\noindent Equation \eqref{DerDensity} implies that density $\rho$  goes to zero at late times exponentially fast; the RHS is always positive for $-1\le w_\infty <1$.
Equation \eqref{DerBeta} shows that $\beta$ can increase if  $w_{\infty} > 1/3$. 
From \eqref{shiftL}, at late times we have
\bea
	w_{eff}(t) = -1 + \frac{\rho}{\Lambda}(1+w_{\infty}) \label{EffectiveEoS}
\eea
Since $\rho$ goes to zero at late times, this means that 
we can write $w_{eff}(t) = -1 + \Omega(t)$ and 
\begin{align}
\frac{\dot{\Omega}}{\Omega} = \frac{\dot{\rho}}{\rho} = -3H_{\infty}(1+w_{\infty})\Bigg(\frac{\coth{\beta}-\frac{1}{3}\tanh{\beta}}{\coth{\beta}-w_{\infty}\tanh{\beta}}\Bigg) \label{Condition}
\end{align}
The quantity within the big parenthesis on the RHS is strictly greater than 1 in the range $1\ge w_\infty > 1/3$ for any non-zero value of $\beta$.
This means that in this range,
\bea
	\frac{\dot{\Omega}}{\Omega}  <  -4H_{\infty} \ \ {\rm or} \ \ 
	\Delta  >  0
\eea
where $\Delta = -	\dot{\Omega}/\Omega-4H_{\infty}$. This is the same condition on $\beta$ increase that we derived from the first order equations in our general perturbative analysis, but let us emphasize that in this subsection we have reached this conclusion under more special circumstances -- the total equation of state is now assumed to be of the specific form \eqref{shiftL} arising from an explicit cosmological constant in the system. While less general, this has the advantage that we are able to relate the $\beta$-increase condition to a specific demand on the (asymptotic) EoS of the {\em rest} of the matter, namely that $w_\infty > 1/3$ (under some other minor assumptions). For $\Lambda$-$w$ system with constant $w$, we have checked these claims by explicit numerical evolution in the full non-linear dipole cosmology equations for $w \gtrsim 0.35 > 1/3$. We expect that with some scanning in the space of initial conditions and better numerics, one should be able to come closer to the $w=1/3$ limit, but we will not try to do this systematically here.  

We close this part by reminding the reader that in this section we have focused on the late time behavior of Universes with a cosmological constant. However, let us emphasize that as discussed in section \ref{sec:dipole-const-w-Eos}, for constant $w$ dipole models without $\Lambda$, with large enough $w$ we can again get tilt-growing cosmologies. 

\subsection{Cosmological Constant vs Flowing Dark Energy}\label{sec:flowingdark}

In the previous sub-section, we exploited the fact that the presence of a cosmological constant is related to the total effective EoS $\rightarrow -1$ at late times. In this sub-section, we will show that the converse statement is not true: the condition that the total effective EoS $\rightarrow -1$, can {\em not} always be translated into the statement that there is an explicit cosmological constant. The technical reason behind this, is easy to see already at the level of FLRW models with a decreasing EoS vs those with an explicit cosmological constant. So we will phrase the discussion in that context, but the idea generalizes to dipole cosmology as well.

Let us start by observing some simple facts. The full set of equations for an FLRW system with general EoS $w_{eff}(t)$ (but without an explicit $\Lambda$) is
\bea
	\Big(\frac{\dot{a}}{a}\Big)^{2} - \frac{k}{a^{2}} = \frac{\rho_{eff}}{3} \label{FlrwWtL0} \\
	\dot{\rho_{eff}} + 3H\rho_{eff}(1+w_{eff}(t)) = 0
\label{FlrwWtL}
\eea
Consider the replacement: 
\bea
\rho_{eff} &=& \rho + \Lambda, \label{shift1} \\ 
w_{eff} &=& -1+\frac{\rho}{(\rho+\Lambda)}(1+w(t)) \label{shift2}
\eea
The resulting equations are: 
\bea
		\Big(\frac{\dot{a}}{a}\Big)^{2}-\frac{k}{a^{2}} = \frac{\rho}{3} + \frac{\Lambda}{3}\\
		\dot{\rho} + 3 H \rho (1+w(t)) = 0
\label{FlrwWthL}	
\eea
These are precisely the equations of an FLRW system, but {\em with} an explicit cosmological constant. Despite the increased complexity of the dipole cosmology system, the shifts \eqref{shift1} and \eqref{shift2} work also in the dipole cosmology system, and their sole effect is the generation of a cosmological constant in the equations of motion. Note in particular that the second equation \eqref{shift2} has the same form as that in \eqref{shiftL}. A crucial fact here is that the choice of $\Lambda$ is arbitrary. This is manifested by the fact that the system above remains of the same form under the replacement
\bea
	\rho \rightarrow  \rho-c,\;\; \Lambda \rightarrow \Lambda + c,\;\; w(t) \rightarrow -1 + \frac{\rho}{\rho-c}(1+w(t)) \label{symmetry}
\eea
where $c$ is arbitrary. Again, this is an invariance\footnote{We call it an invariance (of the system of ODEs) and not a symmetry, to emphasize the fact that not just the fields, but also the ``parameters'' of the theory ($\Lambda$ and the chosen $w(t)$) are getting replaced.} of the dipole cosmology system (written in terms of EoS $w$ and not pressure $p$) with a cosmological constant, and not just of FLRW. A key point is that the freedom to shift $\Lambda$ is inseparable from the shift in $\rho$ and the redefinition in $w$. 

With these preliminaries, the main point is straightforward to make. When we add a  genuine cosmological constant $\Lambda$ to a system whose asymptotic equation of state goes to $w_\infty$ (where $w_\infty$ is strictly greater than $-1$), then we found that the late-time value of $\rho$ tends to zero (and therefore $\rho_{eff} \rightarrow \Lambda$) and $w_{eff} \rightarrow -1$. The crucial point is that when translated to the $(\rho_{eff}, w_{eff})$ language, the late time behavior of all these solutions is such that $\rho_{eff} \rightarrow \Lambda$, a {\em fixed} constant. However, in a system without a cosmological constant, where $w_{eff}\rightarrow -1$, we need not have the property that $\rho_{eff} \rightarrow$ a {\em fixed} constant at late times for all the solutions. Indeed, this is what we found in the explicit examples we discussed in the last section (eg., dipole ``\lcdm'' model) where a glance at the plots reveals that the late time value of $\rho$ is {\em not} the $same$ constant in all the curves\footnote{Note that even in the FLRW limit, one can see a hint of this from the fact that the second term in \eqref{FlrwWtL} is qualitatively different when $w_{eff}=-1$ as opposed to when $w_{eff}> -1$.}. This means that one cannot do a $solution$-$independent$ redefinition and re-interpret the asymptotic value of $\rho$ as a cosmological constant. Note however, that the perturbation theory that we did in the previous subsections remains intact! The zeroth order perturbation equations behave {\em as though} the asymptotic $\rho$ has the interpretation of a cosmological constant around that $particular$ solution\footnote{It is interesting that late time behavior of $w_{eff}(t) \rightarrow -1$ solutions has this curious feature (not just in dipole cosmology, but also in FLRW). This seems to have not been emphasized in the literature before. 
}.

In the context of dipole cosmology, we will refer to those cases with increasing $\beta$ with $w_{eff} \rightarrow -1$ that cannot be re-interpreted as a cosmological constant (plus tilted fluid), as flowing dark energy. The explicit numerical examples we studied in the previous section belong to this class, but we have demonstrated in a previous subsection of this section that explicit cosmological constants can also result in increasing flows as long as the fluid EoS is sufficiently stiff. 

\subsection{Transient Flows}\label{sec:away-from-far-future}

So far in this section we have analyzed late time (ie., after cosmic acceleration has set in) behavior of specific dipole cosmologies and discussed when $\beta$ can grow. However, for viable model building, it may also be interesting to explore dipole cosmologies at earlier epochs. It may happen that $\dot{\beta}$ can change sign.  It may also be interesting to understand if a flow that was created at an earlier epoch can sustain to the present era.

In what follows we present two such examples where $\beta$ grows for a period of time and then it goes to zero at late times. The first example is a system described by a $w(t)$ defined via the matter and radiation of flat \lcdm\ cosmology -- this generalizes our earlier dipole ``\lcdm" model where our focus was only on dust (ie., after radiation has decoupled). We determine the radiation and dust (pressureless matter) history of flat \lcdm\ to define $w_{\text{\tiny{DR}}}(t)$ via
\begin{equation}\label{w-DR}
    w_{\text{\tiny{DR}}}(t)=\frac{p_{\text{\tiny{R}}}(t)+p_{\text{\tiny{D}}}(t)}{\rho_{\text{\tiny{R}}}(t)+\rho_{\text{\tiny{D}}}(t)}=\frac{\rho_{\text{\tiny{R}}}(t)/3}{\rho_{\text{\tiny{R}}}(t)+\rho_{\text{\tiny{D}}}(t)}.
\end{equation}
We then couple a tilted fluid with this equation of state, and a positive cosmological constant, to dipole cosmology. At late times  $w_\infty \rightarrow 0$ for this system, so we do not expect $\beta$ to increase at late times according to our previous general discussion. Explicit numerical evolution confirms this. But we also find that at intermediate times, this system can exhibit transient increase in $\beta$, which we plot in the figures.

\begin{figure}[H]
\centering
\subfloat[\label{fig:LExpWt}]{\includegraphics[width = 7.5 cm]{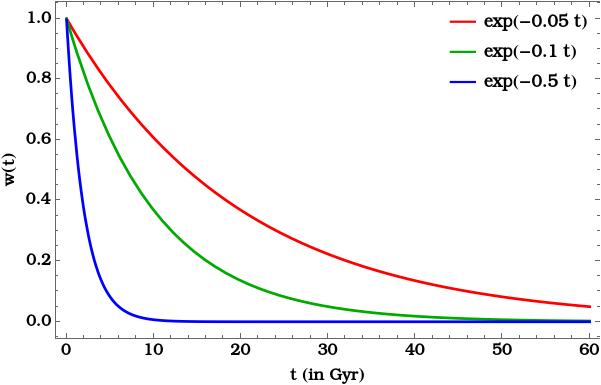}}\hfill
\subfloat[\label{fig:ExpWb}]{\includegraphics[width = 7.5 cm]{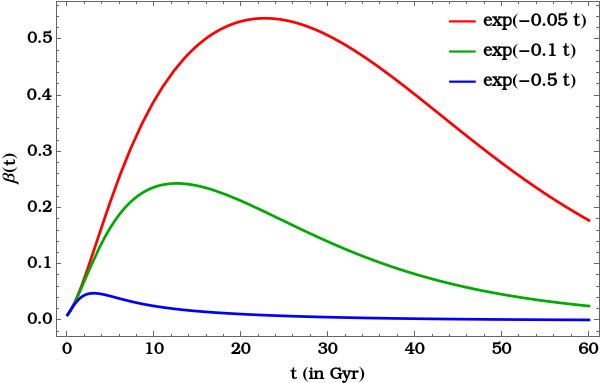}}\\
\subfloat[\label{fig:LLCDMWt}]{\includegraphics[width = 7.5 cm]{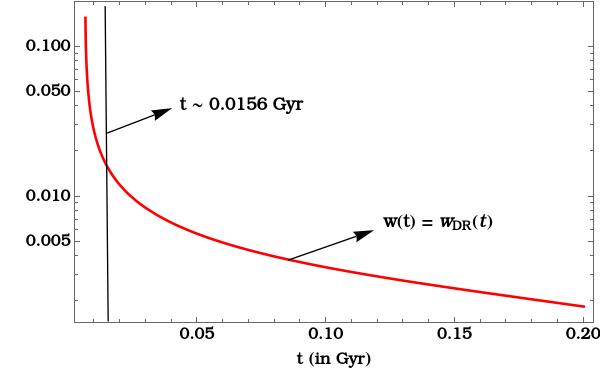}}\hfill
\subfloat[\label{fig:LMWb}]{\includegraphics[width = 7.5 cm]{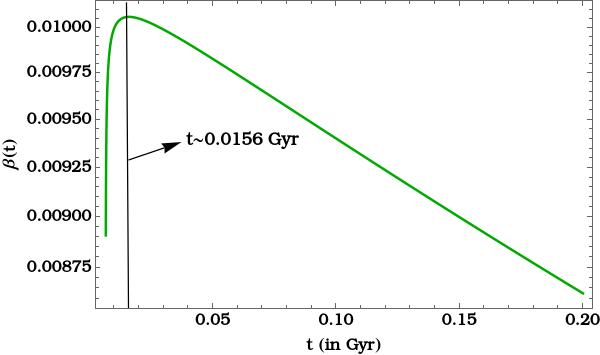}}
\caption{$w(t)$ and $\beta$ for two types of cases, $w=e^{-\alpha t}$ (upper) and $w=w_{\text{\tiny{DR}}}$ given in \eqref{w-DR} (lower) plots. The parameter $\Lambda$ has been kept fixed at 0.0109. We evolved all the systems for initial values $X_{in}=Y_{in}=1$, $\rho_{0} = 0.01$ and $\beta = 0.01$. For upper figures, the initial time $t=0.01$ Gyr, and for the lower figures, it is $t=0.0065$ Gyr. As we see there exists some initial values for which $\beta$ has a bump, it grows and then drops. 
}
\label{fig:Wbeta-bump}
\end{figure}

\begin{figure}[H]
\centering
\subfloat[\label{fig:LExpWt2}]{\includegraphics[width = 8.0 cm]{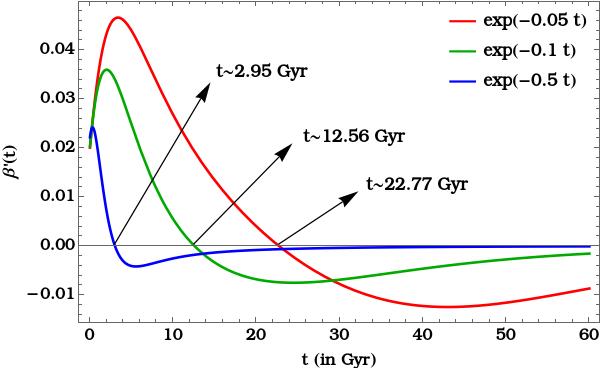}}\hfill
\subfloat[\label{fig:ExpWb2}]{\includegraphics[width = 8.0 cm]{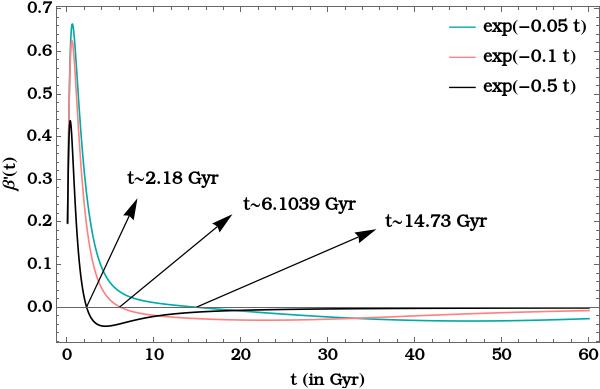}}
\caption{Evolution of $\dot{\beta}(t)$ for two different initial conditions for $w(t) = \exp{(-\alpha t)}$. For the left hand side $X_{in} = Y_{in} = 1.0$ and for the right hand side $X_{in} = Y_{in} = 0.1$ with the other initial parameters remaining the same as mentioned in the caption of (\ref{fig:Wbeta-bump}). We see the zeros of the derivatives (peaks of $\beta$s) appear at different places as the initial conditions change for the same value of $\alpha$.}
\label{fig:WDbeta-bump}
\end{figure}

The second example is when $w(t)$ is  exponentially falling off to zero, $w(t)= e^{-\alpha t}$, together again with a positive cosmological constant.  The  $a(t)$ and $\beta(t)$ plots for the dipole cosmologies in these two cases are given in Figures \ref{fig:Wbeta-bump} and \ref{fig:WDbeta-bump}. 

\section{Concluding Remarks and Outlook}
\label{sec:discussion}

Our goal was to provide the simplest theoretical framework for cosmological model building which accommodates a (not entirely kinematic) cosmic dipole. To this end, we first  phrased a ``dipole cosmological principle'' which led us directly to a specific class of tilted Bianchi cosmologies  (see \cite{Ellis-Maartens-McCallum--Book, King, Ellis-lectures} for context). We worked out the field equations for dipole cosmologies and studied analytically and numerically these equations for some representative classes of cosmic fluids. 

Among other things, we were particularly interested in exploring the evolution of the tilt and whether it is possible to keep $\beta$ sizeable as the universe expands. We found that for accelerating Universes, like what we have  during (quasi-de Sitter slow-roll) inflation or the present day dark energy domination, the tilt can grow in time even though the geometry isotropizes exponentially fast. We also found other  simple examples of cosmologies with growing flows: a single component cosmic fluid with a constant equation of state $w>1/3$ can have a growing tilt, and this can happen with or without a positive cosmological constant. These results are compatible with those in \cite{Coley-2006}. 

The key message is that tilt (which parameterizes a bulk flow in the energy momentum filling the Universe), need not track shear (which parameterizes the anisotropy in the background metric). Explicitly, from \eqref{shear-EoM} and \eqref{EoM-H-sigma-d} we learn
\begin{equation}\label{sigma-die-off}
    \dot{\sigma}=\sigma(-3H+\frac{2A_0}{X}\tanh\beta),
\end{equation}
while for a system described by $w(t)\neq -1$,
\begin{equation}\label{beta-growth-w(t)}
\dot{\beta}\big(\coth\beta-w \tanh \beta\big)=-\frac{\dot{w}}{1+w}+(3w-1)H-\frac23\sigma-\frac{2w A_0\tanh\beta}{X}.
\end{equation}
For an expanding cosmology and assuming $\beta>0$, the last term in \eqref{sigma-die-off} is subdominant as $X$ is increasing due to expansion and $\tanh\beta\leq 1$. So the dynamics of shear is dominated by the cosmic friction $-3H\sigma$ term, whereas for the tilt it is dominated by $(3w-1)H-\dot{w}/(1+w)$ in \eqref{beta-growth-w(t)}. The comparison between the two shows that the former always has a definite negative sign while the latter can change sign -- this is the basis of the different dynamics of $\sigma$ and $\beta$. The above discussions can set the stage for a refinement of the cosmic (no-)hair theorem, in tilted cosmologies \cite{letter}. We have a background ``geometric hair'' (the shear $\sigma$) and a ``cosmic fluid hair'' (the tilt $\beta$). These two kinds of hair are in general nonlinearly coupled, and yet one is able to make statements about late time dynamics: the cosmic fluid hair can grow while the geometric hair is cut short. Our analysis \cite{letter} shows that FLRW geometries with a positive cosmological constant may be unstable against tilt perturbations, even though they may be stable against shear perturbations (as expected from cosmic no-hair theorem). 

In a Universe with a cosmological constant and a tilted fluid with EoS $w(t)$ (which we assume is $>-1$), we found that tilt can increase at late times only if the asymptotic value of $w(t)$ is sufficiently stiff ($ > 1/3$). But we also found examples in which the tilt increases at late times in an accelerating dipole cosmology where the $total $ EoS $w_{eff}(t) \rightarrow -1$, without any explicit cosmological constant. We called this ``flowing dark energy'', to distinguish it from an explicit cosmological constant with a flow. Since the late time EoS of the matter (ie., excluding $\Lambda$) in flat \lcdm\ has an asymptotic EoS that tends to that of dust (ie., $w_\infty=0$), this suggests that the flowing dark energy scenario may be of phenomenological interest in understanding late time flows. In the FLRW setting, these EoS are perfectly consistent with reasonable energy conditions, and it is plausible that this is true in dipole cosmology as well.  A related result we noted was that in a Universe with a cosmological constant, where the rest of the fluid has a flow, we can have intermediate phases of increasing $\beta$. 

While our results were in the specific context of the simplest (Copernican) dipole cosmology, the fact that flow velocities can increase in accelerating Universes is likely to be more generally true. The robustness of our results strongly suggest that variations of these statements should hold in more general (homogeneous or inhomogeneous) tilted cosmologies. It will be interesting if these late time flows can (at least partially) explain away Hubble tension and other cosmic tensions. It has been suggested that the variation of the Hubble constant across the sky is about $10 \%$, and that an isotropic Hubble tension may need a serious rethinking \cite{Rameez}. In fact it has specifically been noted that there is a dipole in the $H_0$ measured from various cosmic sources (eg. \cite{Beyond-FLRW-review} for an detailed review). These and the numerous other tentative hints of non-kinematic cosmic dipoles that we mentioned in the introduction, make dipole flows an observational possibility. The results in this paper provide a theoretical reason for giving them a fair hearing.

Before we conclude however, let us also mention various caveats in our results here. Firstly and most importantly, our assumptions have forced us to have only one independent component in the dipole flow. In other words, except for the possibility of a cosmological constant, all fluids in a dipole cosmology flow at the same velocity. This seems artificial, since the fluids familiar from flat LCDM (ordinary matter, dark matter and radiation) have all largely decoupled from each other. We evaded this problem in our approach by considering time dependent total equations of state as opposed to component fluids with constant equations of state. This is useful for getting general lessons, but it will be necessary to go beyond this if we want to do detailed phenomenology. If we drop the assumption of homogeneity, we will have to deal with PDEs instead of ODEs. 
Perhaps perturbation theory around an FLRW background will be useful -- at late times it may be useful to track the non-linear evolutions of certain specific modes.  A second observation, again one that was forced on us by symmetry, is that we were working with a Universe whose natural isotropic limit is a $k=-1$ FLRW Universe. It may be useful to consider systems where the the $k=0$ limit is accessible while the dipole flow is non-vanishing. This can be done while retaining the homogeneity assumption, and we will have more to say about questions of this kind in another paper \cite{KM}. 

More generally, it is desirable to construct more realistic dipole cosmology models. Such constructions may allow for a tunable non-kinematical CMB dipole (say, in the pre-recombination era). They may also be useful for realizing bulk flows that can yield realistic dipole anisotropy in various sources 
or in our local cluster. 
  
A final issue we will raise, has less to do with our dipole cosmology paradigm as much as the origin of the CMB dipole in the heliocentric frame. If the dipole is non-kinematical, we will have to explain why the dipole flow velocity of our galaxy as inferred from the CMB is comparable to the virialized velocities of galaxies observed in clusters. The two have at least superficially, very different dynamical origins. This can be viewed as a fine-tuning problem for the non-kinematic dipole idea, and was first noted in \cite{Krishnan:2022fzz}.

We have devoted the last few paragraphs to playing devil's advocate to the dipole cosmology paradigm. Despite this, we feel that these ideas are worthy of exploration for the reasons we have detailed throughout the paper. Of particular interest is the fact that direct evidence for the acceleration of the Universe comes from late time observations, while there is a glaring $\sim 5 \sigma$ observational  difficulty in late time cosmology in the form of $H_0$ and other tensions. Together with the challenges to the construction of an accelerating Universe in UV-complete settings like string theory \cite{Swampland-papers} 
it seems evident that (whichever way the chips may eventually fall) late time cosmology is in need of better understanding. Given the current status of cosmic tensions, it has become conceivable that we need a paradigm shift in cosmology. The $\beta$-growth that we noticed in this paper is a possible loophole that can move us away from asymptotic homogeneity and isotropy in an accelerating Universe. Perhaps it is time to replace the Cosmological Principle with the Dipole Cosmological Principle, or something even weaker. 


\section*{Acknowledgments}

We thank Rajeev Jain, Roya Mohayaee and Subir Sarkar for discussions and Eoin \'O.~Colg\'ain for discussions and comments on a version of the draft. CK thanks the participants of the Cosmological Principle  Workshop (25-28 October 2021) at APCTP, Pohang, for questions and comments on a talk based on some of this material.

\appendix 

\section{Alternative Perturbation Theory}\label{sec:appendix-pert}

In this Appendix, we will present late time perturbation theory in the $X, Y$ variables instead of the $a, b$ variables. As before, we take our EoS to be $w(t) = -1 + \epsilon\,\Omega(t)$, where we will view $\Omega(t)$ as a decreasing function that limits to zero at late times. By plugging in 
\bea
X(t)=X_0(t) + \epsilon X_1(t),  \ \ Y(t)= Y_0(t) + \epsilon Y_1(t),  \ \ \beta(t)=\beta_0(t)+ \epsilon \beta_1(t), \ \rho(t)= \rho_0(t)+\epsilon \rho_1(t), \nonumber 
\eea
into the Dipole Friedmann equations \eqref{FW1}, \eqref{FW2}, \eqref{Con1}, \eqref{Con2}, we get the zeroth order equations:
\bea
   \Big(\frac{\dot{X}_{0}}{X_{0}}-\frac{\Dot{Y}_{0}}{Y_{0}}\Big) = 0 \\
   2\frac{\dot{X}_{0}}{X_{0}}\frac{\Dot{Y}_{0}}{Y_{0}} + \Big(\frac{\Dot{Y}_{0}}{Y_{0}}\Big)^{2} -\frac{3}{X_{0}^{2}} = \rho_{0}\\
   \frac{\dot{\rho}_{0}}{\rho_{0}} = 0
\eea
These equations can be solved straightforwardly, and analytic de sitter-like solutions can be written down as before. The first order equations are:
\begin{align}
\Delta_x-\Delta_y&= \rho_{0}\Omega(t)\sinh{\beta_{0}}\cosh{\beta_{0}}X_{0}\\
2\Big(\frac{\dot{X}_{0}}{X_{0}}\Big)(\Delta_{x} + 2\Delta_{y}) +\frac{6 X_{1}}{X_{0}^{3}} &= \rho_{1} + \rho_{0}\Omega(t)\sinh^{2}{\beta_{0}}\\
\frac{\Dot{\rho}_{1}}{\rho_{0}}+ \Omega(t)\Big(3\frac{\dot{X}_{0}}{X_{0}} + \tanh{\beta}_{0}\Dot{\beta}_{0}-\frac{2 \tanh{\beta_{0}}}{X_{0}}\Big) &= 0 \\
\Dot{\Omega}(t) + \Omega(t)\Big(\coth{\beta}_{0}\Dot{\beta_{0}} + \frac{\dot{X}_{0}}{X_{0}} \Big) - \frac{\Dot{\rho}_{1}}{\rho_{0}} &= 0
\end{align}
We have introduced the first order corrections to the Hubble rates along different directions $\Delta_{x}$ and $\Delta_{y}$ defined via 
\begin{align}
    \Delta_{x} \equiv \frac{X_{1}}{X_{0}}\Big(\frac{\dot{X}_{1}}{X_{1}}-\frac{\dot{X}_{0}}{X_{0}}\Big); \;\;   \Delta_{y} \equiv \frac{Y_{1}}{Y_{0}}\Big(\frac{\dot{Y}_{1}}{Y_{1}}-\frac{\dot{Y}_{0}}{Y_{0}}\Big) 
\end{align}
The shear (which only get contributions at first order) is in fact controlled directly by these quantities
\bea
\sigma = (\Delta_x-\Delta_y). 
\eea
Along with the shear, the tilt also gets contributions only  at first order. Decoupling the last two of the first order equations above we get simple equations for the tilt and density perturbation:
\bea
\dot{\beta}_{0} & =& \frac{1}{2}\tanh{2\beta_{0}}\left(-\frac{\dot{\Omega}}{\Omega}-4\frac{\dot{X}_{0}}{X_{0}}+2\frac{\tanh{\beta_{0}}}{X_{0}}\right) \\
\frac{\dot{\rho}_{1}}{\rho_{0}} &=& -\frac{1}{2}\left(3\frac{\dot{X}_{0}}{X_{0}}(\coth{\beta_{0}}-\frac{1}{3}\tanh{\beta_{0}})-\frac{\dot{\Omega}}{\Omega}\tanh{\beta_{0}}-\frac{2}{X_{0}}\right)
\eea
For some choices of $\Omega(t)$, $\beta_0$ can be solved analytically at late times, and even in general, it is easily determined numerically. Both the shear equation above (the first of the first order equations) and the tilt equation allow us to find the asymptotic solutions for these quantities easily. 

It is in the form presented in this Appendix that we have matched late time perturbation theory with our exact numerical evolution plots. It is straightforward to reach the form presented in section \ref{sec:5} from the one presented here, via the variable change to $a, b$ from $X, Y$. 

\section{Asymptotic Behavior of Tilted Energy}

In this Appendix, we will argue that the asymptotic value of the Hubble constant tends to $\sqrt{\Lambda/3}$ in many cases of interest. This is unsurprising, but it is instructive to see some of the details of this, in the context of dipole cosmology ---- see eg. below \eqref{shiftL} for a discussion.

The relevant equation of motion is
\begin{equation}
H^{2} =  \frac{1}{9}\sigma^{2} + \frac{A_{0}^{2}}{a^{2}}e^{-4b}  + \frac{\rho}{3} + \frac{1}{3}(\rho + p)\sinh^{2}{\beta} + \frac{\Lambda}{3}\label{FieldEq}
\end{equation}
Our goal will be to argue fairly generically, that all terms on the RHS except $\Lambda$, die down at late times. 
Assuming the fluid EoS saturates to $w_{\infty}$ asymptotically, the conservation equation for $\rho$ becomes \eqref{laterho}
\begin{equation}
\dot{\rho} + \rho(1+w_{\infty})\Big(3H + \tanh{\beta}\dot{\beta}\Big) = 0 \label{DensityDer}
\end{equation} 
$H(t)$ is always greater than zero, and $-1 < w_{\infty} < 1$ is the domain of EoS that we are interested in. 
We will consider $\beta$ that is always positive. If $\dot{\beta}$ is positive, this guarantees that $\rho$ dies down at late times exponentially. Even if $\dot{\beta} < 0$ at late times, since we will only be concerned with $\beta$ that is bounded below by zero, $\dot{\beta}$ must go to zero from below at late times and $\beta$ must saturate to some non-negative value. This again ensures that $\rho$ dies down at late times, and we can ignore the third term on the RHS in \eqref{FieldEq}. We also expect the first and second terms to die down because shear is decreasing and the Universe is expanding. We have checked this in many examples, numerically. In fact for shear, it is easy to see that the demand that it vanishes asymptotically from \eqref{shear-EoM} is precisely equivalent to the forth term on the RHS of \eqref{FieldEq} dying down at late times. So we have a self-consistent scenario where $H$ tends to $\sqrt{\Lambda/3}$, if the term $\propto (\rho + p)\sinh^{2}{\beta} \rightarrow 0$ at late times. 

This quantity $(\rho + p)\sinh^{2}{\beta}$ is interesting for a few different reasons, so we will elaborate on it a bit before demonstrating that it indeed vanishes at late times.  The energy momentum tensor of a tilted fluid \eqref{tilted-EM-up-down}, is of the form of a usual isotropic perfect fluid plus a tilt piece. Note that the contribution to the $00$ component of the energy momentum tensor from the tilted part is nothing but ${\cal E}:=(\rho+p)\sinh^2\beta$. We will call it the tilted energy. As we noted, ${\cal E}$ features on the right-hand-side of \eqref{EoM-H-sigma-c} and captures the effects of the tilt in $H^2$. 

When $\beta$ is decreasing, since $\rho$ is decreasing anyway, it should be clear that ${\cal E}$ also decreases with time. As discussed in the main text, for constant $w$ the $\beta$-evolution is controlled by the equation
\begin{equation}
\dot{\beta}(\coth{\beta}-w\tanh{\beta}) = (3w-1)H - \frac{2}{3}\sigma - \frac{2A_{0}w\tanh{\beta}}{a(t)}e^{-2b(t)}
\end{equation}   
In the context of our present discussion, at late times this becomes
\begin{equation}
\dot{\beta}(\coth{\beta}-w_{\infty}\tanh{\beta}) = (3w_{\infty}-1)H \label{betaDer}
\end{equation}
For any finite $\beta > 0$, we have $\coth{\beta} > \tanh{\beta}$. So for $|w_{\infty}| < 1$, the term $(\coth{\beta}-w_{\infty}\tanh{\beta}) > 0$. Hence when $w_{\infty} < 1/3$, we have $\dot{\beta} < 0$ and trivially, ${\cal E}$ decreases at late times. 

Interestingly, the conclusion that ${\cal E}$ is vanishing at late times is in fact valid even when $w_\infty < 1/3$ and $\beta$ is increasing. To see this, note that the desired condition is
\begin{equation}
\frac{d}{dt}\log{{\cal E}} \equiv \frac{\dot{\rho}+\dot{p}}{\rho + p} + 2\coth{\beta}\dot{\beta} < 0 
\end{equation}
Under the assumptions we are working with, the late time conservation equations take the form
\begin{subequations}
\begin{eqnarray}
\dot{\rho}+3H(\rho+p)=&-(\rho+p)\tanh\beta\dot{\beta} \label{Con1-Appen} \\
\dot{p}+H(\rho+p)=& -(\rho+p)\coth{\beta}\dot{\beta} \label{Con2-Appen}
\end{eqnarray}
\end{subequations}
where $p\equiv w_\infty \rho$. 
Using these expressions as well as \eqref{betaDer}, we find that 
\begin{subequations}\begin{align}
\frac{1}{{\cal E}}\frac{d {\cal E}}{dt} &= \frac{\dot{\rho}+\dot{p}}{\rho + p} + 2\coth{\beta}\dot{\beta} \\
 &= -4H - \tanh{\beta}\dot{\beta} + \coth{\beta}\dot{\beta} \\
 &= -H\left(4 - (3w_{\infty}-1)\frac{(\coth{\beta}-\tanh{\beta})}{\coth{\beta}-w_{\infty}\tanh{\beta}}\right) 
\end{align}\end{subequations}
The final expression in parenthesis on the RHS is manifestly positive for $-1 < w_\infty <1$. In particular, this is true irrespective of whether $w_\infty > 1/3$ or not.  This establishes that the tilted energy indeed dies down at late times, establishing that the asymptotic value of $H$ in these cases is $\sqrt{\Lambda/3}$. 


\begin{thebibliography}{99}

\bibitem{Copernicus}
https://en.wikipedia.org/wiki/De\_revolutionibus\_orbium\_coelestium

%

\bibitem{Spradlin}
M.~Spradlin, A.~Strominger and A.~Volovich,
``Les Houches lectures on de Sitter space,''
[arXiv:hep-th/0110007 [hep-th]].

\bibitem{WeinbergOldBook}
S.~Weinberg,
``Gravitation and Cosmology: Principles and Applications of the General Theory of Relativity,'' John Wiley and Sons, 1972;
``Cosmology,'' Oxford University press, 2008.

\bibitem{Ellis-Maartens-McCallum--Book}
Ellis, G. F. R., Maartens, R. and MacCallum, M. A. H. (2012). Relativistic cosmology. Cambridge University Press. https://doi.org/10.1017/CBO9781139014403


  
\bibitem{crisis}
L.~Verde, T.~Treu and A.~G.~Riess,
``Tensions between the Early and the Late Universe,''
Nature Astron. \textbf{3}, 891
[arXiv:1907.10625 [astro-ph.CO]].

\bibitem{Intro0} E.~Di Valentino, L.~A.~Anchordoqui, O.~Akarsu, Y.~Ali-Haimoud, L.~Amendola, N.~Arendse, M.~Asgari, M.~Ballardini, S.~Basilakos and E.~Battistelli, \textit{et al.} ``Cosmology intertwined II: The Hubble constant tension,'' Astropart. Phys. \textbf{131}, 102605 (2021)
[arXiv:2008.11284 [astro-ph.CO]].

 
\bibitem{Eleonora-et-al-review-1}
E.~Di Valentino, O.~Mena, S.~Pan, L.~Visinelli, W.~Yang, A.~Melchiorri, D.~F.~Mota, A.~G.~Riess and J.~Silk,
``In the realm of the Hubble tension\textemdash{}a review of solutions,''
Class. Quant. Grav. \textbf{38} (2021) no.15, 153001
[arXiv:2103.01183 [astro-ph.CO]].

\bibitem{SNOWMASS-2022}
E.~Abdalla, G.~Franco Abell\'an, A.~Aboubrahim, A.~Agnello, O.~Akarsu, Y.~Akrami, G.~Alestas, D.~Aloni, L.~Amendola and L.~A.~Anchordoqui, \textit{et al.}
``Cosmology intertwined: A review of the particle physics, astrophysics, and cosmology associated with the cosmological tensions and anomalies,''
JHEAp \textbf{34} (2022), 49-211
doi:10.1016/j.jheap.2022.04.002
[arXiv:2203.06142 [astro-ph.CO]].

\bibitem{Planck-2018} N.~Aghanim \textit{et al.} [Planck],
``Planck 2018 results. VI. Cosmological parameters,''
Astron. Astrophys. \textbf{641} (2020), A6
[erratum: Astron. Astrophys. \textbf{652} (2021), C4]
[arXiv:1807.06209 [astro-ph.CO]].

\bibitem{Eleonora-combined-analysis}
E.~Di Valentino,
``A combined analysis of the $H_0$ late time direct measurements and the impact on the Dark Energy sector,''
Mon. Not. Roy. Astron. Soc. \textbf{502} (2021) no.2, 2065-2073
[arXiv:2011.00246 [astro-ph.CO]].


\bibitem{Anjan}
C.~Krishnan, E.~\'O.~Colg\'ain, Ruchika, A.~A.~Sen, M.~M.~Sheikh-Jabbari and T.~Yang,
``Is there an early Universe solution to Hubble tension?,''
Phys. Rev. D \textbf{102}, no.10, 103525 (2020), 
[arXiv:2002.06044 [astro-ph.CO]].

C.~Krishnan, E.~\'O.~Colg\'ain, M.~M.~Sheikh-Jabbari and T.~Yang,
``Running Hubble Tension and a H0 Diagnostic,''
Phys. Rev. D \textbf{103} (2021) no.10, 103509
[arXiv:2011.02858 [astro-ph.CO]].


C.~Krishnan, R.~Mohayaee, E.~\'O.~Colg\'ain, M.~M.~Sheikh-Jabbari and L.~Yin,
``Does Hubble tension signal a breakdown in FLRW cosmology?,''
Class. Quant. Grav. \textbf{38} (2021) no.18, 184001
[arXiv:2105.09790 [astro-ph.CO]].




\bibitem{Beyond-FLRW-review}
P.~K.~Aluri, P.~Cea, P.~Chingangbam, M.~C.~Chu, R.~G.~Clowes, D.~Hutsem\'ekers, J.~P.~Kochappan, A.~Krasi\'nski, A.~M.~Lopez and L.~Liu, \textit{et al.}
``Is the Observable Universe Consistent with the Cosmological Principle?,''
Class. Quant. Grav. \textbf{40} (2023) no.9, 094001, 
[arXiv:2207.05765 [astro-ph.CO]].






\bibitem{SubirEtc} 
N.~J.~Secrest, S.~von Hausegger, M.~Rameez, R.~Mohayaee, S.~Sarkar and J.~Colin,
``A Test of the Cosmological Principle with Quasars,''
Astrophys. J. Lett. \textbf{908} (2021) no.2, L51
[arXiv:2009.14826 [astro-ph.CO]];



J.~Colin, R.~Mohayaee, M.~Rameez and S.~Sarkar,
``Evidence for anisotropy of cosmic acceleration,''
Astron. Astrophys. \textbf{631} (2019), L13
[arXiv:1808.04597 [astro-ph.CO]].

R.~Mohayaee, M.~Rameez and S.~Sarkar,
``The impact of peculiar velocities on supernova cosmology,''
[arXiv:2003.10420 [astro-ph.CO]].

A.~K.~Singal,
``Peculiar motion of Solar system from the Hubble diagram of supernovae Ia and its implications for cosmology,'' Mon. Not. Roy. Astron. Soc. \textbf{515} (2022) no.4, 5969-5980, 
[arXiv:2106.11968 [astro-ph.CO]].

N.~Horstmann, Y.~Pietschke and D.~J.~Schwarz,
``Inference of the cosmic rest-frame from supernovae Ia,''
Astron. Astrophys. \textbf{668} (2022), A34, 
[arXiv:2111.03055 [astro-ph.CO]].

\bibitem{Rameez}
M.~Rameez and S.~Sarkar,
``Is there really a Hubble tension?,''
Class. Quant. Grav. \textbf{38} (2021) no.15, 154005
[arXiv:1911.06456 [astro-ph.CO]].


\bibitem{SNe-hemisphere-anomaly}
C.~Krishnan, R.~Mohayaee, E.~\'O.~Colg\'ain, M.~M.~Sheikh-Jabbari and L.~Yin,
``Hints of FLRW breakdown from supernovae,''
Phys. Rev. D \textbf{105} (2022) no.6, 063514
[arXiv:2106.02532 [astro-ph.CO]].

\bibitem{Subir2}
V.~V.~Makarov and N.~J.~Secrest,
``Testing the Cosmological Principle: Astrometric Limits on Systemic Motion of Quasars at Different Cosmological Epochs,''
Astrophys. J. Lett. \textbf{927} (2022) no.1, L4
[arXiv:2202.07536 [astro-ph.GA]];

N.~Secrest, S.~von Hausegger, M.~Rameez, R.~Mohayaee and S.~Sarkar,
``A Challenge to the Standard Cosmological Model,''
[arXiv:2206.05624 [astro-ph.CO]].

A.~K.~Singal,
``Solar system peculiar motion from the Hubble diagram of quasars and testing the cosmological principle,''
Mon. Not. Roy. Astron. Soc. \textbf{511} (2022) no.2, 1819-1829
[arXiv:2107.09390 [astro-ph.CO]].

\bibitem{QSO-hemisphere-anomaly}

O.~Luongo, M.~Muccino, E.~\'O.~Colg\'ain, M.~M.~Sheikh-Jabbari and L.~Yin,
``Larger H0 values in the CMB dipole direction,''
Phys. Rev. D \textbf{105} (2022) no.10, 103510
[arXiv:2108.13228 [astro-ph.CO]];

E.~\'O.~Colg\'ain, M.~M.~Sheikh-Jabbari, R.~Solomon, G.~Bargiacchi, S.~Capozziello, M.~G.~Dainotti and D.~Stojkovic,
``Revealing Intrinsic Flat $\Lambda$CDM Biases with Standardizable Candles,''
Phys. Rev. D \textbf{106} (2022) no.4, L041301, 
[arXiv:2203.10558 [astro-ph.CO]].


\bibitem{Ellis-Baldwin} 
 G. F. R. Ellis and J. E. Baldwin, ``On the Expected Anisotropy of Radio Source Counts,'' Monthly Notices of the Royal Astronomical Society, Volume 206, Issue 2, January 1984.

\bibitem{Saha}
S.~Saha, S.~Shaikh, S.~Mukherjee, T.~Souradeep and B.~D.~Wandelt,
``Bayesian estimation of our local motion from the Planck-2018 CMB temperature map,''
JCAP \textbf{10}, 072 (2021)
doi:10.1088/1475-7516/2021/10/072
[arXiv:2106.07666 [astro-ph.CO]].

\bibitem{Quartin}
P.~d.~Ferreira and M.~Quartin,
``First Constraints on the Intrinsic CMB Dipole and Our Velocity with Doppler and Aberration,''
Phys. Rev. Lett. \textbf{127}, no.10, 101301 (2021)
doi:10.1103/PhysRevLett.127.101301
[arXiv:2011.08385 [astro-ph.CO]].

\bibitem{King}
A.~R.~King and G.~F.~R.~Ellis,
``Tilted homogeneous cosmological models,''
Commun. Math. Phys. \textbf{31} (1973), 209-242

\bibitem{Ellis-lectures}
G.~F.~R.~Ellis and H.~van Elst,
``Cosmological models: Cargese lectures 1998,''
NATO Sci. Ser. C \textbf{541} (1999), 1-116
[arXiv:gr-qc/9812046 [gr-qc]].



\bibitem{Wald-cosmic-no-hair}
R.~M.~Wald,
``Asymptotic behavior of homogeneous cosmological models in the presence of a positive cosmological constant,''
Phys. Rev. D \textbf{28} (1983), 2118-2120.

\bibitem{Stewart-Ellis}
J.~M.~Stewart and G.~F.~R.~Ellis,
``Solutions of Einstein's equations for a fluid which exhibit local rotational symmetry,''
J. Math. Phys. \textbf{9} (1968), 1072-1082.

\bibitem{MacCallum}
G.~F.~R.~Ellis and M.~A.~H.~MacCallum,
``A Class of homogeneous cosmological models,''
Commun. Math. Phys. \textbf{12} (1969), 108-141



\bibitem{Tsagas:2021tqa}
C.~G.~Tsagas,
``The deceleration parameter in \textquoteleft{}tilted\textquoteright{} universes: generalising the Friedmann background,''
Eur. Phys. J. C \textbf{82} (2022) no.6, 521
[arXiv:2112.04313 [gr-qc]].

\bibitem{KM}
E. Ebrahimian, C.~Krishnan, R.~Mondol and M.M. Sheikh-Jabbari, ``Towards Dipole $\Lambda$CDM Cosmology'', to appear.


\bibitem{Coley-2006}
A.~A.~Coley, S.~Hervik and W.~C.~Lim,
``Fluid observers and tilting cosmology,''
Class. Quant. Grav. \textbf{23} (2006), 3573-3591
[arXiv:gr-qc/0605128 [gr-qc]].

\bibitem{letter}
C.~Krishnan, R.~Mondol and M.~M.~Sheikh-Jabbari,
``A Tilt Instability in the Cosmological Principle,''
[arXiv:2211.08093 [astro-ph.CO]].

\bibitem{Krishnan:2022fzz}
C.~Krishnan and R.~Mondol,
``$H_0$ as a Universal FLRW Diagnostic,''
[arXiv:2201.13384 [astro-ph.CO]].

\bibitem{Swampland-papers}
G.~Obied, H.~Ooguri, L.~Spodyneiko and C.~Vafa,
``De Sitter Space and the Swampland,''
[arXiv:1806.08362 [hep-th]].

S.~K.~Garg and C.~Krishnan,
JHEP \textbf{11}, 075 (2019)
[arXiv:1807.05193 [hep-th]].

H.~Ooguri, E.~Palti, G.~Shiu and C.~Vafa,
``Distance and de Sitter Conjectures on the Swampland,''
Phys. Lett. B \textbf{788}, 180-184 (2019)
[arXiv:1810.05506 [hep-th]].

S.~K.~Garg, C.~Krishnan and M.~Zaid Zaz,
``Bounds on Slow Roll at the Boundary of the Landscape,''
JHEP \textbf{03}, 029 (2019)
[arXiv:1810.09406 [hep-th]].

\end{thebibliography}
\end{document}